\documentclass{aastex631}

\shorttitle{Electron cosmic rays}
\shortauthors{Yang et al.}
\graphicspath{{./}{figures/}}

\newcommand{\comment}[1]{}

\newcommand{\changetext}[1]{#1}
\newcommand{\changetextmath}[1]{{\textcolor{brown}{ #1}}}

\usepackage{amsmath}
\usepackage{appendix}

\begin{document}

\title{Compton scattering of electrons in the intergalactic medium}

\author[0000-0001-8849-4866]{Yuanyuan Yang}
\affiliation{Center for Cosmology and Astroparticle Physics,
191 West Woodruff Avenue,
Columbus, Ohio 43210, USA}
\affiliation{Department of Astronomy,
140 West 18th Avenue,
Columbus, Ohio 43210, USA}
\affiliation{Khoury College of Computer Science, Northeastern University,
440 Huntington Ave, Boston, Massachusetts 02115, USA}

\author[0000-0001-8290-5417]{Heyang Long}
\affiliation{Center for Cosmology and Astroparticle Physics,
191 West Woodruff Avenue,
Columbus, Ohio 43210, USA}
\affiliation{Department of Physics,
191 West Woodruff Avenue,
Columbus, Ohio 43210, USA}

\author[0000-0002-2951-4932]{Christopher M. Hirata}
\affiliation{Center for Cosmology and Astroparticle Physics,
191 West Woodruff Avenue,
Columbus, Ohio 43210, USA}
\affiliation{Department of Astronomy,
140 West 18th Avenue,
Columbus, Ohio 43210, USA}
\affiliation{Department of Physics,
191 West Woodruff Avenue,
Columbus, Ohio 43210, USA}

\begin{abstract}
This paper investigates the distribution and implications of cosmic ray electrons within the intergalactic medium (IGM). Utilizing a synthesis model of the extragalactic background, we evolve the spectrum of Compton-included cosmic rays.
The energy density distribution of cosmic ray electrons peaks at redshift $z \approx2$, and peaks in the $\sim$MeV range. 
The fractional contribution of cosmic ray pressure to the general IGM pressure progressively increases toward lower redshift. At mean density, the ratio of cosmic ray electron to thermal pressure in the IGM $ P_{\rm CRe} / P_{\rm th}$ is 0.3\% at $z=2$, rising to 1.0\% at $z=1$, and 1.8\% at $z=0.1$ \changetext{(considering only the cosmic rays produced locally by Compton scattering)}.
We compute the linear Landau damping rate of plasma oscillations in the IGM caused by the $\sim$MeV cosmic ray electrons, and find it to be of order $\sim 10^{-6}\,\rm s^{-1}$ for wavenumbers $1.2\lesssim ck/\omega_{\rm p}\lesssim 5$ at $z=2$ and mean density (where $\omega_{\rm p}$ is the plasma frequency).
This strongly affects the fate of TeV $e^+e^-$ pair beams produced by blazars, which are potentially unstable to oblique instabilities involving plasma oscillations with wavenumber $ck/\omega_{\rm p}\approx\sec\theta$ ($\theta$ being the angle between the beam and wave vector). Linear Landau damping is at least thousands of times faster than either pair beam instability growth or collisional effects; it thus turns off the pair beam instability except for modes with very small $\theta$ ($ck/\omega_{\rm p}\rightarrow 1$, where linear Landau damping is kinematically suppressed).
This leaves open the question of whether the pair beam instability is turned off entirely, or can still proceed via the small-$\theta$ modes.
\end{abstract}

\section{Introduction} \label{sec:intro}

Cosmic ray electrons are ubiquitous in the Universe. They are present locally near Earth (see \citealt{2000ApJ...532..653B, 2011PhRvL.106t1101A, 2019PhRvL.122j1101A} for measurements down to GeV energies), and can now be measured down to low ($\sim$few MeV) energies beyond the heliopause \citep{2013Sci...341..150S, 2016ApJ...831...18C}. Within the Milky Way's interstellar medium (ISM), cosmic ray electrons are responsible for the principal sources of emission in radio (synchrotron emission) and contribute significantly to the diffuse gamma rays via inverse Compton and bremsstrahlung emission \citep[see, e.g.,][]{2012ApJ...750....3A, 2013MNRAS.436.2127O, 2016ApJS..223...26A} and are modeled in standard tools to predict radio and gamma ray emission \citep[e.g.][]{2018NPPP..297..129O}. On even larger scales, cosmic ray leptons are visible via radio haloes and relics in galaxy clusters (see \citealt{2008SSRv..134...93F, 2012A&ARv..20...54F, 2019SSRv..215...16V} for reviews). 

By volume, most of the Universe is filled with the intergalactic medium (IGM), and much remains unknown regarding the role of non-thermal components such as leptonic and hadronic cosmic rays and magnetic fields in the low-density IGM phases. Cosmic rays are generally understood to be accelerated in collisionless shocks \citep{1977ICRC...11..132A, 1978MNRAS.182..147B, 1978ApJ...221L..29B}. Structure formation shocks, such as those at filaments \changetext{\citep[e.g.][]{2004ApJ...617..281K, 2021MNRAS.505.4178V}}, provide a natural site for accelerating cosmic ray electrons, and modern cosmological simulations of cosmic rays contain a source term based on fits to microphysical shock simulations \citep{2017MNRAS.465.4500P, 2023MNRAS.519..548B}. Galactic winds could also advect cosmic rays into the IGM, or provide additional shocks for (re-)acceleration. Alternatively, cosmic rays could diffuse into the IGM, although the length scale over which they can diffuse is uncertain \citep{2015MNRAS.448L..20L}. Nevertheless, most of the volume of the IGM is in voids that have been subjected at most to very weak shocks, e.g., from relaxation following reionization \citep{2004MNRAS.348..753S, 2018MNRAS.474.2173H}. Diffusive acceleration in weaker shocks produces a much softer spectrum of cosmic rays: the phase space density declines with momentum as $f(p)\propto p^{-4/(1-{\cal M}^{-2})}$ instead of $f(p)\propto p^{-4}$ (where $p$ is the momentum and ${\cal M}$ is the Mach number\footnote{This is simply a consequence of the smaller compression ratios for weaker shocks. We have used $\gamma=5/3$ and the weak magnetic field limit, as appropriate for the ionized IGM. See, e.g., the textbook derivation in \citet{2011piim.book.....D}, \S36.2.}), whereas to produce an electron with a collisional momentum loss rate $(d\ln p/dt)_{\rm coll}=H(z)$ one must reach quasi-relativistic momenta, e.g., $p\approx 1.4m_ec$ at $z=2$ and mean density.\footnote{This is based on the standard collisional energy loss formulae for electrons, as detailed later in this paper.} Furthermore, this spectrum only exists if non-thermal seed electrons are present on which diffusive shock acceleration can act: this ``injection problem'' for electrons, and the circumstances under which it is solved, are a subject of current research \citep[e.g.][]{2010PhRvL.104r1102A, 2011ApJ...733...63R, 2021ApJ...919...97K, 2022ApJ...932...86S}. Electrons may also be present as secondary cosmic rays, however low Mach number shocks in plasmas with weak magnetic fields may not always efficiently accelerate protons or He ions \citep{2018ApJ...864..105H}.

It is therefore of interest to investigate sources of cosmic ray electrons that could apply in the full volume of the IGM. Gamma rays produced by sources such as active galactic nuclei (AGN) are neutral and thus can propagate freely into the IGM. While the Universe is mostly transparent to $\sim$MeV gamma rays, a small percentage of them will interact in the IGM via Compton scattering:
\begin{equation}
\gamma + e^-({\rm th})\rightarrow \gamma+e^-({\rm cr}),
\end{equation}
where ``th'' and ``cr'' denote thermal and cosmic ray electrons, respectively. \changetext{We focus on the electrons because the cross section for Compton scattering by heavier particles such as protons is lower than for electrons, by approximately 6 order of magnitude \citep{PhysRev.73.449}. Therefore this ``guaranteed'' source of cosmic rays principally produces electrons. (If proton cosmic rays are present in a given region, they could produce an additional source of electrons and positrons due to Bethe-Heitler pair production in collisions with background photons; or due to charged pion production and decay in collisions with matter.)}
%While it is true that protons, depending on their energy, can interact with photons or gas, leading to processes like Bethe-Heitler pair production, these interactions are less efficient in generating cosmic-ray electrons. As detailed in the work of \cite{PhysRev.73.449}, the efficiency of Compton scattering for protons is significantly lower than for electrons, by approximately 6 order of magnitude.
Compton scattering of the X-ray background is a potential source of heating in the IGM \citep{1999ApJ...517L...9M}, which is most important in low-density regions where photoionization heating is weaker\footnote{After reionization, a photoionization must be preceded by a recombination; thus it inherits the density-squared dependence of recombination.} \citep{2012ApJ...746..125H}. Compton scattering of X-rays will produce non-relativistic recoil electrons that thermalize their energy in a cosmologically short time, but gamma ray photons in the $\sim$few MeV range can produce quasi-relativistic electrons ($p/m_ec$ of order unity). At IGM densities and low redshift ($z\lesssim 5$), these quasi-relativistic electrons can survive for cosmologically long timescales, since neither collisional losses nor inverse Compton cooling via the cosmic microwave background (CMB: important for much higher energies) are fast. This provides a ``guaranteed'' source of cosmic ray electrons in the IGM. The central calculation in this paper is to take a synthesis model of the extragalactic background and evolve the spectrum of Compton-included cosmic rays in the IGM.

There are several consequences for a ubiquitous cosmic ray electron population everywhere in the IGM. In this paper, we focus our detailed calculations on the most spectacular result: cosmic rays modify the dispersion relation of plasma waves in the IGM. This affects the fate of the TeV pair beams produced by blazars via
\begin{equation}
\gamma({\rm TeV}) + \gamma ({\rm EBL}) \rightarrow e^+ + e^-,
\end{equation}
where ``$\gamma$(TeV)'' denotes a TeV photon emitted by a blazar and ``EBL'' denotes the typically $\sim$eV extragalactic background light. These pair beams can cool via inverse Compton scattering, but it has also been proposed that they drive an instability of plasma oscillations that could disrupt the beam and lead to an additional heating source in the IGM \citep{2012ApJ...752...22B, 2012ApJ...752...23C}. It has been debated whether detuning due to density (and hence plasma frequency) gradients \citep{2013ApJ...770...54M} or transfer of energy to longer-wavelength plasma oscillation modes \citep{2014ApJ...787...49S} stops the instability from growing (but see, e.g., \citealt{2013ApJ...777...49S, 2014ApJ...797..110C, 2018ApJ...859...45S} for counterarguments). Limits on the $\sim 100$ GeV gamma rays produced by inverse Compton process further motivate understanding whether the plasma instability mechanisms can cool pair beams \citep[e.g.][]{2023arXiv230301524B}. We will show in this paper that the $\sim$MeV electrons cause linear Landau damping of plasma oscillations; the damping rates can be thousands of times faster than would occur for a purely thermal plasma. This suppresses the plasma instabilities in most (but not quite all) of the relevant parameters space.

This paper is organized as follows: In Section \ref{sec:con}, we present the formalism of electron cosmic rays. In Section \ref{sec:model}, we discuss the different components of the model. Section \ref{sec:sol} provides a numerical solution. We present the results and contribution of the electron cosmic rays to the IGM pressure, energy density, and plasma oscillation damping rate in Section \ref{sec:res}. Finally, Section~\ref{sec:fin} discusses some of the other potential implications of the cosmic ray electrons that might be explored in future work.

\section{Conventions and conversions}
\label{sec:con}

We want to construct a model to represent the Compton-scattered electron cosmic ray distribution within the Intergalactic Medium (IGM). To start, it is pivotal to parameterize the cosmic ray spectrum. An approach to achieve this is to define $N(E)$ as the number of electrons per unit physical volume per unit energy $E=m_ec^2(\gamma-1)$ [units: cm$^{-3}$ erg$^{-1}$]. Then there is an energy content per baryon in cosmic rays of $\int E N(E)\,dE/n_b$, where $n_b$ is the physical baryon density.

\changetext{Since we are focused here on building a model of ``guaranteed'' sources of cosmic rays in the diffuse regions of the IGM that have not fallen into a structure formation shock (e.g., filament shock), we have not included shock crossing or diffusion of cosmic rays from high-density regions. If a shock passes over a region, then this could lead to both new cosmic rays as well as further acceleration of existing cosmic rays (via diffusive shock acceleration). It is unclear whether we should expect diffusion of cosmic rays to be significant in the IGM. If diffusion is significant, it would lead to a more uniform distribution of cosmic rays across the IGM. It would also mean that shock-accelerated particles in filaments, galaxies, or clusters could diffuse into the IGM and contribute to the cosmic ray population in near-mean density regions. Because of this possibility, a calculation of the {\em in situ} produced cosmic rays in the IGM, neglecting shocks and diffusion, provides a lower bound on the cosmic ray population in mean density regions.}
%\changetext{In formulating the governing evolution equation, we have made specific assumptions, notably the neglect of diffusion and the absence of shock crossing. While these assumptions simplify the model, they also introduce potential limitations that merit consideration.}
%\changetext{We acknowledge that Compton scattering is a ubiquitous phenomenon in the IGM, and our model aims to calculate a lower limit of its impact. We recognize that shocks are present in the IGM, but due to the lack of detailed knowledge about their distribution and intensity, their significance in this context remains uncertain. Our approach, therefore, represents a simplified scenario that prioritizes a general overview over detailed interactions in specific locales.}
%\changetext{As for the diffusion of cosmic rays, if this process is significant, it could lead to a more uniform distribution of cosmic rays across the IGM. This would imply that our model, which assumes a mean density ($\rm log_{10} \Delta_0$ = 0), could be representative of a broad range of conditions within the IGM. However, it is crucial to note that the absence of detailed modeling of diffusion and shock crossing in our current framework means that our predictions are more aligned with a general representation rather than a detailed account of localized variations in the cosmic ray distribution.}

Under the assumptions of neglecting diffusion and the absence of a shock crossing, the governing evolution equation can be stated as:
\begin{eqnarray}
\frac{\partial}{\partial t} N(E) = &-&\theta N(E)
+ \frac13 \theta \frac\partial{\partial E} \left[\frac{E(E+2m_ec^2)}{E+m_ec^2}  N(E) \right]
-
\frac{\partial }{\partial E} 
\left[\left.\frac{dE}{dt}\right|_{\rm loss}N(E)\right]
\nonumber \\
&+& n_{e,\rm tot}\int_{2E}^\infty c\beta' \frac{d\sigma_{\rm M}(E|E')}{dE} N(E') dE' + S(E,t),
\label{eq:evol}
\end{eqnarray}
where $\theta = 3H - d\ln\Delta/dt$ is the divergence of the baryon velocity and $\Delta$ is the relative overdensity. The first two terms in this equation account for the dilution of the density and the momentum alteration for adiabatic expansion. When adiabatic expansion induces a stretch in the de Broglie wavelength of a particle by a factor of $\Delta\lambda/\lambda$, the consequent energy decrease is given by $\Delta E = -\Delta\lambda/\lambda \times E(E+2m_ec^2)/(E+m_ec^2)$. The term $(dE/dt)|_{\rm loss}$ represents energy loss rate. Moreover, $\sigma_{\rm M}(E|E')\,dE$ represents the M\o ller scattering cross section for an electron of energy $E'$ to produce a secondary electron of energy between $E$ and $E+dE$. Lately, $S(E,t)$ is defined as the source term.

Within the confines of our computational model, we use the Centimeter-Gram-Second (CGS) unit system, although the outputs can be translated into eV-based units to facilitate more intuitive interpretation. In the CGS framework, the electron mass is represented as $m_ec^2 = 8.187\times 10^{-7}\,$erg. Furthermore, all cross section are expressed in terms of the classical electron radius, denoted by $r_0 = 2.818\times 10^{-13}\,$cm.

Electrons will be described by their kinetic energy, symbolized as $E$. In relation to this kinetic energy, the Lorentz factor and the velocity, when expressed as a fraction of the speed of light, are described as follows:

\begin{equation}
\gamma = \frac{E}{m_ec^2}+1
{\rm ~~~and~~~}
\beta = \frac{\sqrt{E(E+2m_ec^2)}}{E+m_ec^2}.
\end{equation}

We use the cosmological model from \cite{2020A&A...641A...6P}: $H_{\rm 0} = 67.4 \,\rm km\, s^{-1} Mpc^{-1}$, $\Omega_{\rm m} = 0.315$ and $\Omega_{\rm b}\rm h^2= 0.0224$. 

\section{Model ingredients}
\label{sec:model}

\subsection{Density evolution}

Our model for the gas density evolution is a spherical over- (or under-) density in a $\Lambda$CDM background universe. Due to Birkhoff's theorem, a uniform spherical overdensity has a radius that satisfies the usual Friedmann equation \citep{1967ApJ...147..859P}. Writing the radius as $a\xi$, where $a$ is the scale factor and the overdensity is $\Delta = \xi^{-3}$, the second-order Friedmann equation is
\begin{equation}
\frac{d^2}{dt^2}(a\xi) = -\frac{\Omega_mH_0^2}{2(a\xi)^2}
+ \frac12(1-\Omega_m)H_0^2a\xi
\end{equation}
where the first term comes from the matter density and the second from the cosmological constant. It can be transformed into
\begin{equation}
\ddot\xi = -2H\dot\xi + \frac{\Omega_mH_0^2}{2a^3}\left(\xi-\xi^{-2}\right).
\end{equation}
Here $H = d(\ln a)/dt$ is the Hubble rate.
We initialize this equation at high redshift with $\xi = 1-a\delta_{\rm lin}/3$ and $\dot\xi = -aH\delta_{\rm lin}/3$, where $\delta_{\rm lin}$ is the linear overdensity with the growth function scaled out, and evolve it to low redshift with a leapfrog scheme. $\delta_{\rm lin}$ = 0 means mean density, $\delta_{\rm lin} <$ 0 means underdensity, and $\delta_{\rm lin} >$ 0 means overdensity.

\subsection{Source term}

The source term $S(E,t)$ in Eq.~(\ref{eq:evol}) represents the rate of production of Compton-scattered electrons per unit physical volume per unit energy. It can be expressed in terms of the extragalactic background intensity as
\begin{equation}
S(E,t) = n_{e,\rm tot}\int_{E_{\gamma,\rm min}}^\infty
\frac{d\sigma_{\rm KN}(E|E_\gamma)}{dE}
\frac{4\pi J(E_\gamma)}{E_\gamma}
dE_\gamma,
\end{equation}
where $J(E_\gamma)= J_\nu/h$ is the background intensity at photon energy $E_\gamma=h\nu$. For $J_{\nu}$, we use the fiducial (``Q18'') model in \citet{2019MNRAS.484.4174K} and interpolate the EBL specific intensity via a 4-nearest point method. We also use the model in \cite{2012ApJ...746..125H} for comparison in the discussion. The minimum photon energy to produce a recoil electron of energy $E$ is
\begin{equation}
E_{\gamma,\rm min} = \frac12\left[ E + \sqrt{E(E+2m_ec^2)} \right],
\end{equation}
and the Klein-Nishina cross section is
\begin{equation}
\frac{d\sigma_{\rm KN}(E|E_\gamma)}{dE}
= \pi r_0^2 \frac{m_ec^2}{E_\gamma(E_\gamma-E)}\left[
2 + \frac{E^2}{E_\gamma^2} - 2\frac{E}{E_\gamma} - 2\frac{m_ec^2E}{E_\gamma^2} + \frac{(m_ec^2)^2E^2}{E_\gamma^3(E_\gamma-E)}
\right].
\end{equation}

\begin{figure}
\centering
\vspace{-1.2cm}
\includegraphics[width=120mm]{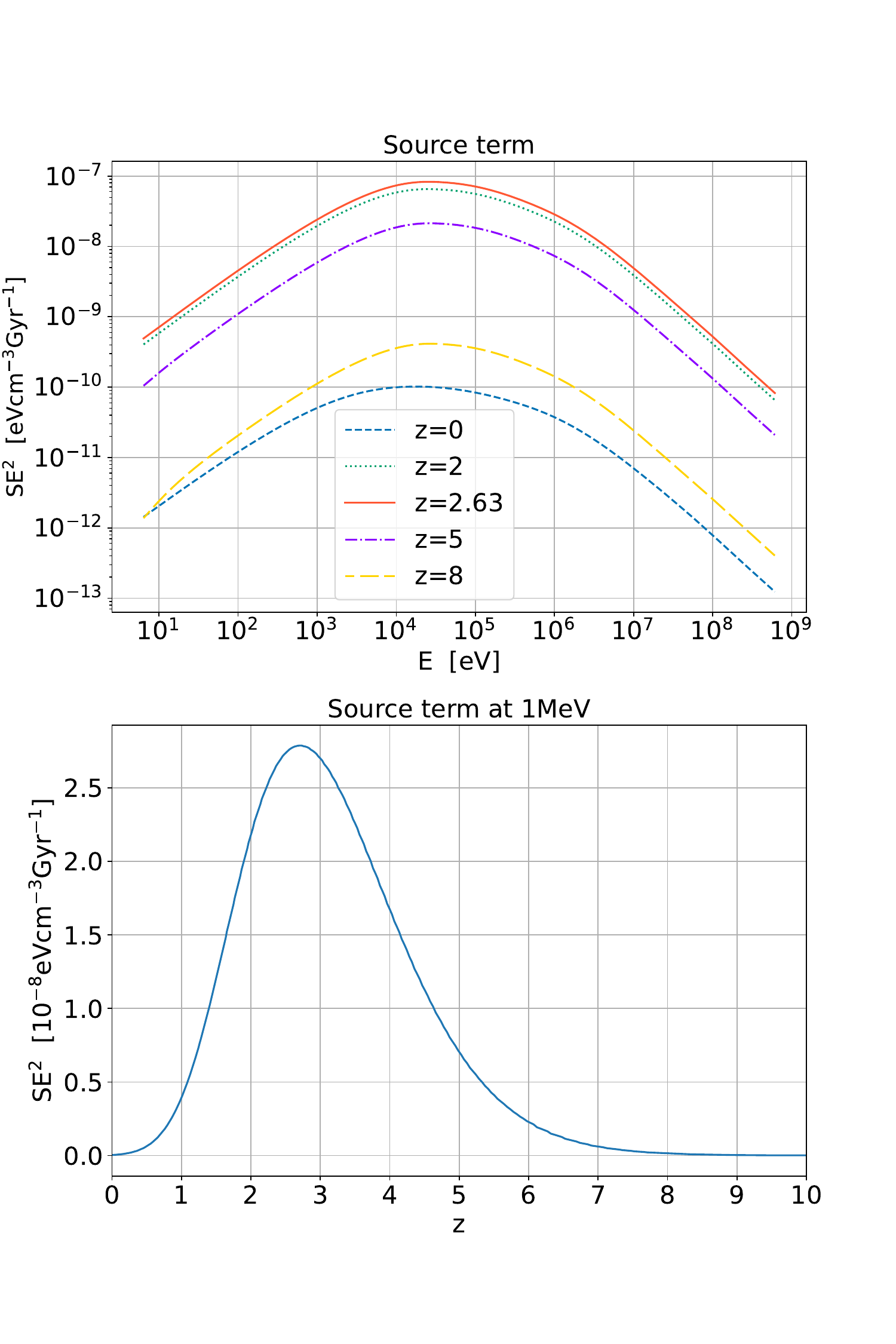}
\vspace{-1.2cm}
\caption{ {\em Top}: The source terms at redshift $z$ = 0, 2, 2.63, 5, and 8 are labeled accordingly. {\em Bottom}: The production rate of 1 MeV electrons at different redshift. These source terms exhibit an increasing trend from $z$ = 15 to $z$ = 3, followed by a decrease from $z$ = 2 to $z$ = 0. Furthermore, they reach their peak around $z \approx$ 2 to 3.
\label{fig:source}}
\centering
\end{figure}

The top panel of figure \ref{fig:source} illustrates the source terms at redshift $z$ = 0, 2, 2.63, 5, and 8. In our comprehensive results, these source terms exhibit an increasing trend from $z$ = 15 to $z$ = 3, followed by a decrease from $z$ = 2 to $z$ = 0, and reach their peak around $z \approx$ 2 to 3. The bottom panel of figure \ref{fig:source} illustrates one example of how source term evolving with redshift for 1 MeV electrons.

The principal source of model uncertainty here --- and generally in this paper --- is the extragalactic background in the $\sim$MeV range. The integrated background near Earth was measured by the {\slshape Solar Maximum Mission} \citep{2000AIPC..510..471W} and {\slshape Compton Gamma Ray Observatory} \citep{2000AIPC..510..467W}. \citet{2019MNRAS.484.4174K} adjusted the parameters of their Type 2 quasar spectral energy distribution (SED) to match the observed background. However, other sources may also contribute to the MeV background, including supernovae and radioactive and cosmic ray-induced processes in the ISM of galaxies; model predictions for these have ranged from a few to $\sim 10\%$ of the observed background \citep{2001PASJ...53..669I, 2005JCAP...04..017S, 2014ApJ...786...40L}. Although it is reasonable to expect that the MeV contribution would follow the overall redshift evolution of both star formation and AGN, peaking at ``Cosmic Noon'' and then declining toward the present day, unresolved background measurements cannot definitively establish whether this is the case. Finally, \citet{2019MNRAS.484.4174K} present a ``point estimate,'' and $\sim$MeV backgrounds a factor of a few lower would also provide an acceptable fit.

\subsection{Energy losses by relativistic electrons}
\begin{figure}
\centering
\includegraphics[width=\textwidth]{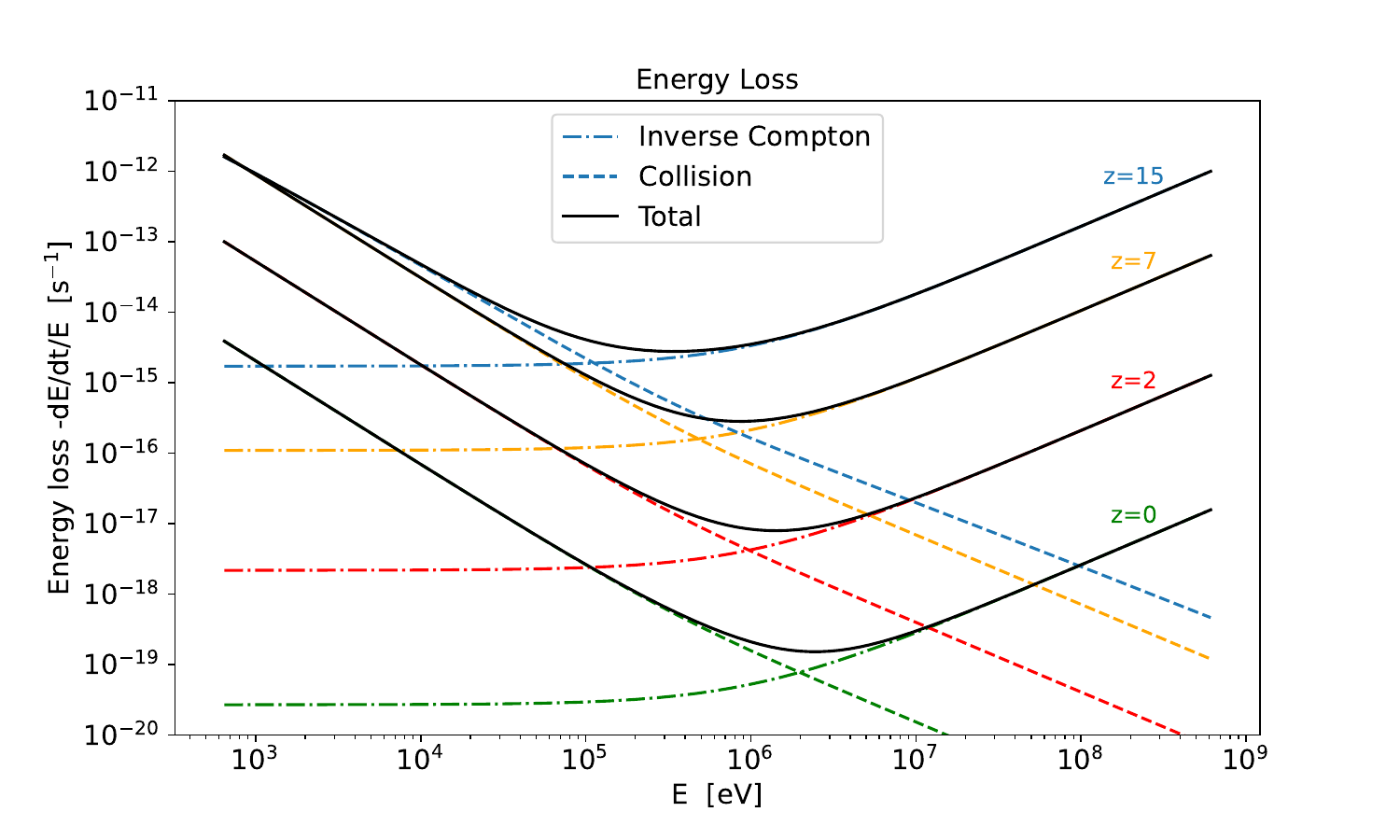}
\caption{\label{fig:loss}The energy loss due to inverse-Compton cooling (dot-dashed line), collision (dashed line), and the total energy loss (solid line) at the mean density for redshift $z=15$, 7, 2, and 0. As the redshift decreases, the energy loss rate also decreases. At lower electron energies ($\leq$ MeV), the dominant energy loss mechanism is collisional loss, whereas at higher electron energy levels, inverse-Compton cooling dominates the overall energy loss. Note the small change in slope at low $E$ from $z=15$ to $z=7$ due to reionization.}
\end{figure}

Electrons of the $\sim$MeV energy range in the intergalactic medium lose energy by two major mechanisms: inverse Compton cooling, and collisional losses (which include both free electrons in the plasma as well as collisions with bound electrons in H$^0$, He$^0$, and He$^+$).

The inverse Compton loss can be computed using standard formulae. The expression of \citet[Eqs.~16,20]{1965PhRv..137.1306J} gives, in the Thomson scattering limit $k_{\rm B}T_{\rm CMB}\gamma \ll m_ec^2$ where we may keep only the leading term in the infinite sum,
\begin{equation}
-\left\langle \frac{d\gamma}{dt} \right\rangle = \frac{32}9 \pi r_0^2 \frac{ a_{\rm R}T_{\rm CMB}^4}{m_ec} \beta^2\gamma^2,
\end{equation}
where $r_0$ is the classical electron radius and $a_{\rm R}$ is the radiation constant.
Written in terms of energy, this is
\begin{equation}
\left.\frac{dE}{dt} \right|_{\rm inv.~Compton} = -\frac{32\pi r_0^2 a_{\rm R}T_{\rm CMB}^4}{9m_ec}  \left( 2 + \frac{E}{m_ec^2}\right)E.
\label{eq:dEdtCompton}
\end{equation}

The energy loss due to collisions with free electrons in an ionized plasma can be computed using the standard formula \citep{1954PhRv...93...38R, 1954ARNPS...4..315U}:
\begin{equation}
\left.\frac{dE}{dt} \right|_{\rm coll.,~free} = -\frac{2\pi n_er_0^2m_ec^3}{\beta} \left[ \ln\frac{E^2}{I^2} + \ln \left(1 + \frac\tau 2\right) + F^- - \delta \right],
\label{eq:BB}
\end{equation}
where $n_e$ is the electron density. The high momentum transfer part of the energy loss, computed using the \citet{1932AnP...406..531M} scattering formula, is
\begin{equation}
F^- = (1-\beta^2) \left[ 1 + \frac{\tau^2}8 - (2\tau+1)\ln 2 \right],
\end{equation}
and $\tau = E/m_ec^2$. For a plasma, the mean excitation energy $I$ is the plasma frequency $\omega_p$:
\begin{equation}
I = \hbar\omega_p, ~~~\omega_p = c\sqrt{4\pi r_0 n_e}.
\end{equation}
The density correction is given by the theory of \citet{1956PhRv..103.1202F,1963ARNPS..13....1F}, with $\Im[-1/\epsilon] = \frac\pi2\omega_p \delta_{\rm D}(\omega-\omega_p)$ and $\ell = \omega_p\beta\gamma$:
\begin{equation}
\delta = 2\ln\gamma - \beta^2.
\end{equation}
This combines to give
\begin{equation}
\left.\frac{dE}{dt} \right|_{\rm coll.,~free} = -\frac{2\pi n_er_0^2m_ec^3}{\beta} \left\{ \ln\frac{E^2}{4\pi \hbar^2 c^2 r_0n_e} + \ln \left(1 + \frac\tau 2\right) + (1-\beta^2) \left[ 1 + \frac{\tau^2}8 - (2\tau+1)\ln 2 \right] - 2\ln\gamma + \beta^2 \right\}.
\label{eq:dEdtCollision}
\end{equation}
Note the presence of the usual Coulomb logarithm factor, as generally occurs in energy loss calculations.

If atoms with bound electrons (H$^0$, He$^0$, or He$^+$) are present, then we should also take into account collisional losses with these species. The collisional loss rate is given by a modified version of Eq.~(\ref{eq:BB}) without a density correction:
\begin{equation}
\left.\frac{dE}{dt} \right|_{\rm coll.,~bound} = -\frac{2\pi r_0^2m_ec^3}{\beta} \sum_{{\rm species}~X} N_{e,X}n_{X}\left[ \ln\frac{E^2}{I_X^2} + \ln \left(1 + \frac\tau 2\right) + F^- \right],
\end{equation}
where $n_X$ is the number density of species $X$; $N_{e,X}$ is the number of bound electrons (1 for H$^0$ or He$^+$, or 2 for He$^0$); and $I_X$ is the geometric mean excitation energy for species $X$ (15.0 eV for H$^0$; 42.3 eV for He$^0$; and 59.9 eV for He$^+$).\footnote{For the single-electron species H$^0$ and He$^+$, we use the hydrogenic oscillator strength distribution. For He$^0$, we use the computation of \citet{2014MolPh.112..751S}.}

We assume that hydrogen (H) undergoes complete ionization instantaneously at $z$ = 8, and helium (He) ionized into He$^+$ completely at $z$ = 8 and experiences complete ionization instantaneously at $z$ = 3. While the ionization of H and He is a graduation processes, we simplify it by assuming instantaneous ionization. This assumption is reasonable because H is fully ionized around $z$ = 6, when cosmic rays are still not abundant. Similarly, although He reionization happens in the presence of abundant cosmic rays, the contribution of it to the overall electron population is only a few percent. Consequently, the simplicity of the instant ionization model does not significantly differ from more complex models.

Figure \ref{fig:loss} displays the energy loss due to inverse-Compton cooling, collisions, and the total energy loss at mean density for redshift $z$ = 0, 2, 7, and 15. The energy loss rate demonstrates a decreasing trend as the redshift decreases. Notably, for electrons in lower energy levels ($\leq$ MeV), collisional loss stands out as the primary energy loss mechanism. Conversely, for electrons in higher energy levels, inverse-Compton cooling plays a dominant role in the overall energy loss.

\changetext{If there is an intergalactic magnetic field, then synchrotron radiation provides an additional possible energy loss mechanism. However, this mechanism is small compared to inverse Compton scattering of the CMB. Even for an equipartition field, the ratio is $ n k T_{\rm gas} / (a T_{\rm CMB}^4)$, which is around $5.1\times 10^{-7}$ at $z=2$. This means for any realistic field strength, the loss by synchrotron is negligible.
Similarly, we expect synchrotron self-Compton (SSC) processes to be negligible, because the ratio of SSC to synchrotron radiation energy density would be of order $\sigma_{\rm T}n_{e,\rm rel} ct\gamma^2$ if a population of relativistic electrons of number density $n_{e,\rm rel}$ can scatter the synchrotron photons for a duration of time $t$. (Here $\sigma_{\rm T}=\frac83\pi r_0^2$ is the Thomson cross section.) For our case we find $\sigma_{\rm T}n_{e,\rm rel} ct\gamma^2\sim 10^{-10}$.}

\subsection{Secondary electrons}

The M\o ller scattering differential cross section to produce a secondary electron of energy $E$ from a primary of energy $E'$ is (see, e.g., \citealt{1982spre.reptQ....B}, Eq.~2.15):
\begin{equation}
\frac{d\sigma_{\rm M}(E|E')}{dE}
= \frac{2\pi r_0^2 m_ec^2(m_ec^2+E')^2}{E'(2m_ec^2+E')E^2}
\left[
1 + \frac{E^2}{(E'-E)^2} + \frac{E^2}{(m_ec^2+E')^2}
- \frac{m_ec^2(m_ec^2+2E')E}{(m_ec^2+E')^2(E'-E)}
\right]
\end{equation}
with limits $E<E'/2$ since we consider the ``secondary'' electron to be the one with lower energy (electrons are identical particles so this is a labeling convention).

\section{Numerical solution}
\label{sec:sol}

Equation~(\ref{eq:evol}) is a partial differential equation for $N(E,t)$ with two independent variables, $E$ and $t$. As redshift $z$ is more commonly used in cosmology, we can convert Equation~(\ref{eq:evol}) to $z$ format by
\begin{equation}
    \frac{\partial}{\partial z} N(E, z) = \frac{dt}{dz} \frac{\partial N(E, t)}{\partial t} = -\frac{1}{(1+z) H(z)} \frac{\partial N(E, t)}{\partial t} .
\end{equation}
This can be re-written in the form
\begin{equation}
    \frac{\partial}{\partial z} N(E, z) = A_1(z) N(E, z) +  \frac{\partial }{\partial E}[A_2(E,z)N(E, z)] + \int_{2E}^{\infty} A_3(E, E',z) N(E', z) dE' + S_z(E,z),
\label{eq:nez}
\end{equation}
where the coefficients $A_1$, $A_2$, $A_3$, and $S_z$ have the indicated dependences and are given by
\begin{eqnarray}
    A_1(z) &=& \frac{\theta}{(1+z) H(z)}, \nonumber \\
    A_2(E,z) &=& -\frac{1}{(1+z) H(z)}\left[ \frac13 \theta \frac{E(E+2m_ec^2)}{E+m_ec^2} - \left.\frac{dE}{dt}\right|_{\rm loss} \right], \nonumber \\
    A_3(E,E',z) &=& -\frac{1}{(1+z) H(z)} n_{e,\rm tot}\ c\beta' \frac{d\sigma_{\rm M}(E|E')}{dE},~~{\rm and} \nonumber \\
    S_z(E,z) &=& -\frac{S(E,t)}{(1+z) H(z)}.
\end{eqnarray}

To integrate Eq.~(\ref{eq:nez}), we discretize the energy axis into $m$ energy bins with logarithmically spaced energies $\{E_i\}_{i=0}^{m-1}$ and widths $\Delta E_i = E_{i+1/2}-E_{i-1/2}$. We track the number density of cosmic ray electrons $n_i(z) = N(E_i,z)\Delta E_i$ (units: cm$^{-3}$) in each bin, and choose the following form:
\begin{equation}
    \frac{dn_i(z)}{dz} = A_1(z) n_i(z) + A_2(E_{i+{1}/{2}},z)\frac{n_{i+1}(z)}{\Delta E_{i+1}}
    - A_2(E_{i-{1}/{2}},z) \frac{n_{i}(z)}{\Delta E_{i}}  + \sum_{j: E_j \geq 2E_i} A_3(E_i, E_j,z) n_j(z) + S_z\Delta E_i.
    \label{eq:evol_simp}
\end{equation}
There are several possible discretizations of the $A_2$ term: the version using $n_{i+1}(z)$ as chosen here, an alternative using $n_{i-1}(z)$, or a symmetric version using both. This version rigorously enforces that information flows only from higher to lower energy bins, and it is stable for the case of electrons losing energy ($A_2<0$).

Writing the $n_i$ as a length-$m$ column vector ${\boldsymbol n}$, Eq.~(\ref{eq:evol_simp}) can be cast in the form
\begin{equation}
    \frac{\partial \boldsymbol n}{\partial z} = {\bf M}{\boldsymbol n} + {\boldsymbol s},
    \label{eq:evol_general}
\end{equation}
where the source vector ${\boldsymbol s}$ has components $s_i = S_z(E_i,z)\Delta E_i$ and ${\bf M}$ is an $m\times m$ upper-triangular matrix.

We solve this using the backward Euler method:
\begin{equation}
{\boldsymbol n}(z_{j+1}) \approx [{\bf I}_{m\times m} - (z_{j+1}-z_j){\bf M}]^{-1}[{\boldsymbol n}(z_j) + (z_{j+1}-z_j){\boldsymbol s}(z_{j+1})],
\label{eq:back-Euler}
\end{equation}
which allows us to inductively determine ${\boldsymbol n}(z_1)$ from ${\boldsymbol n}(z_0)$; then we can find ${\boldsymbol n}(z_2)$ and so on. Since we integrate forward in time, $z_{j+1}-z_j<0$. We initialize with ${\boldsymbol n}(z_0)=0$ at $z_0=15$.
After finding $n_i(z_j)$, we can divide by $\Delta E_i$ to get $N(E_i, z_j)$.

\section{Results}
\label{sec:res}

\subsection{Implication for energy and pressure balance in the IGM}
\begin{figure}
\centering
\includegraphics[width=\textwidth]{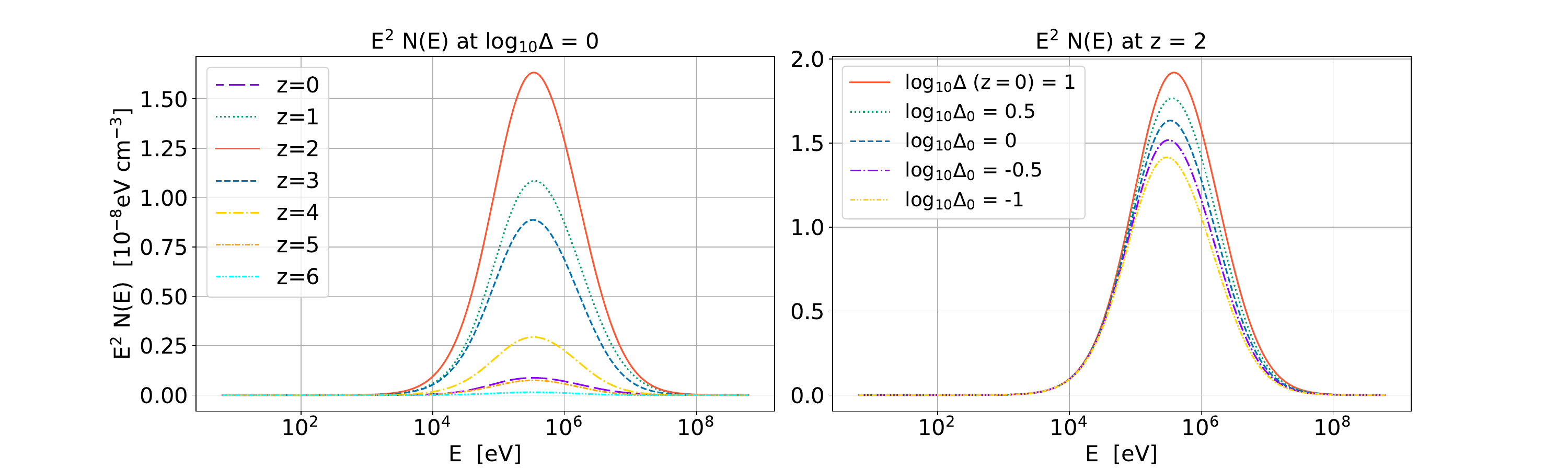}
\caption{Energy density distribution of electrons across different energy bins. {\em Left:} Distribution at mean density for redshift $z = 0 ... 6$, as labeled. It peaks at $z=2$, and then declines, following the overall evolution of AGN emissivity. {\em Right}: Distribution at $z=2$ under different over- or under-densities. The curves are labeled by the value of $\log_{10}\Delta$ at the present epoch (all were closer to mean density at $z=2$). As the matter density increases, the energy density of electrons also rises. In all cases, the energy density of electrons peaks at around the MeV energy level.}
\label{fig:E2N}
\end{figure}

\begin{figure}
\centering
\includegraphics[width=\textwidth]{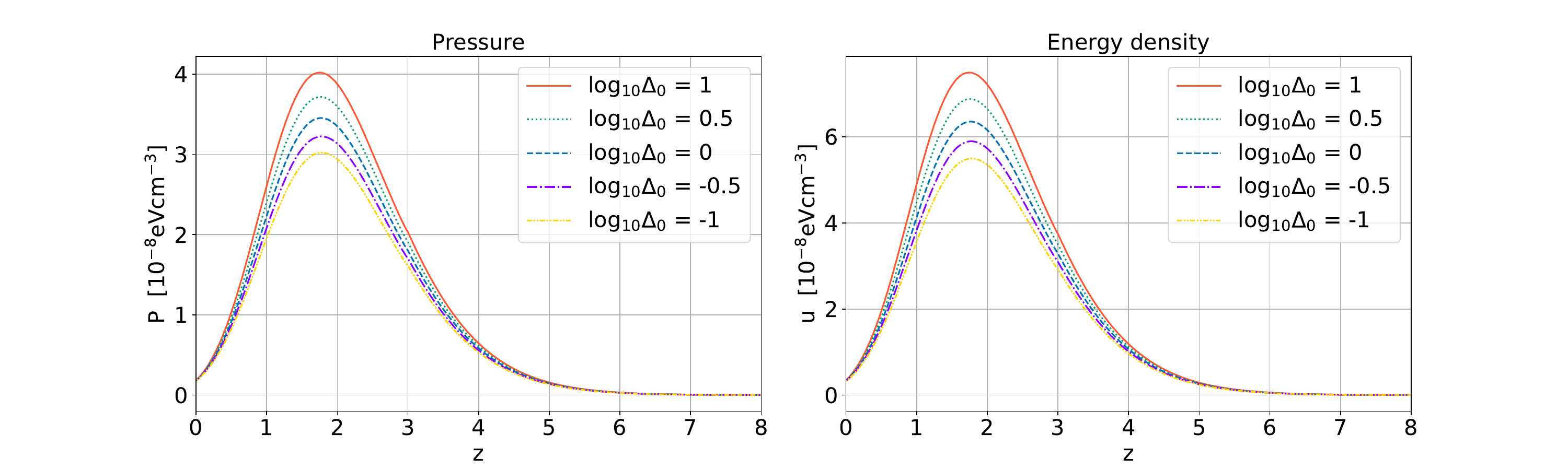}
\caption{The energy density (left) and pressure (right) of electron cosmic rays for different matter over- or under-densities, as labeled. In both plots, we observe an increasing trend from $z$ = 0 to $z \approx$ 2, followed by a decrease as z increases, approaching values close to 0 before $z \approx$ 6. Furthermore, both the energy density and pressure exhibit higher value for higher matter density settings.
\label{fig:utot and P}}
\end{figure}

The total energy density from the electron cosmic rays is
\begin{equation}
    u_{\rm CRe} = \int E N_e(E, z) dE,
\end{equation}
and the pressure is given by the usual special relativistic result:
\begin{equation}
    P_{\rm CRe} = \frac13 \int E \frac{E + 2m_ec^2}{E + m_ec^2} N_e(E, z) dE.
\end{equation}

Figure \ref{fig:E2N} depicts the energy density distribution of electrons across various energy bins. The left panel presents the distribution at mean density for redshifts $z=6$ down through $z=0$.
The electron energies exhibit an increasing trend from $z=6$ to $z=2$, and then decreases toward lower redshift, consistent with the source distribution shown in Figure~\ref{fig:source}.

For the right panel, we focus on the snapshots at $z=2$ and illustrate the distribution at different overdensities. As the matter density increases, the energy density of electrons also rises, which is expected given that the ambient electrons are the targets that are Compton-scattered to produce cosmic rays. At lower energies, the curves converge because both the source and collisional energy losses are proportional to density. Notably, in all cases, the energy density of electrons peaks in the 0.1--1 MeV energy range. This finding aligns with Figure \ref{fig:loss}, which shows that energy losses by electrons are minimized in this range.

Figure \ref{fig:utot and P} displays the energy density (left) and pressure (right) of electron cosmic rays as a function of redshift for various matter over- or under-density configurations. In both plots, we observe an increasing trend until $z\sim 1.7$, followed by a rapid decline toward $z=0$. This pattern is consistent with the findings in Figure \ref{fig:source} and the left plot of Figure \ref{fig:E2N}. We should remember that these are physical energy densities and pressures, and thus the overall expansion of the Universe contributes to the low-$z$ decline: the $\sim$MeV photons emitted at Cosmic Noon are still present as an extragalactic background, but by $z\lesssim 1$ they have been diluted and redshifted. Furthermore, both the energy density and pressure are larger in overdensities, as depicted in the bottom plot of Figure \ref{fig:E2N}.

\begin{figure}
\centering
\includegraphics[height=6in]{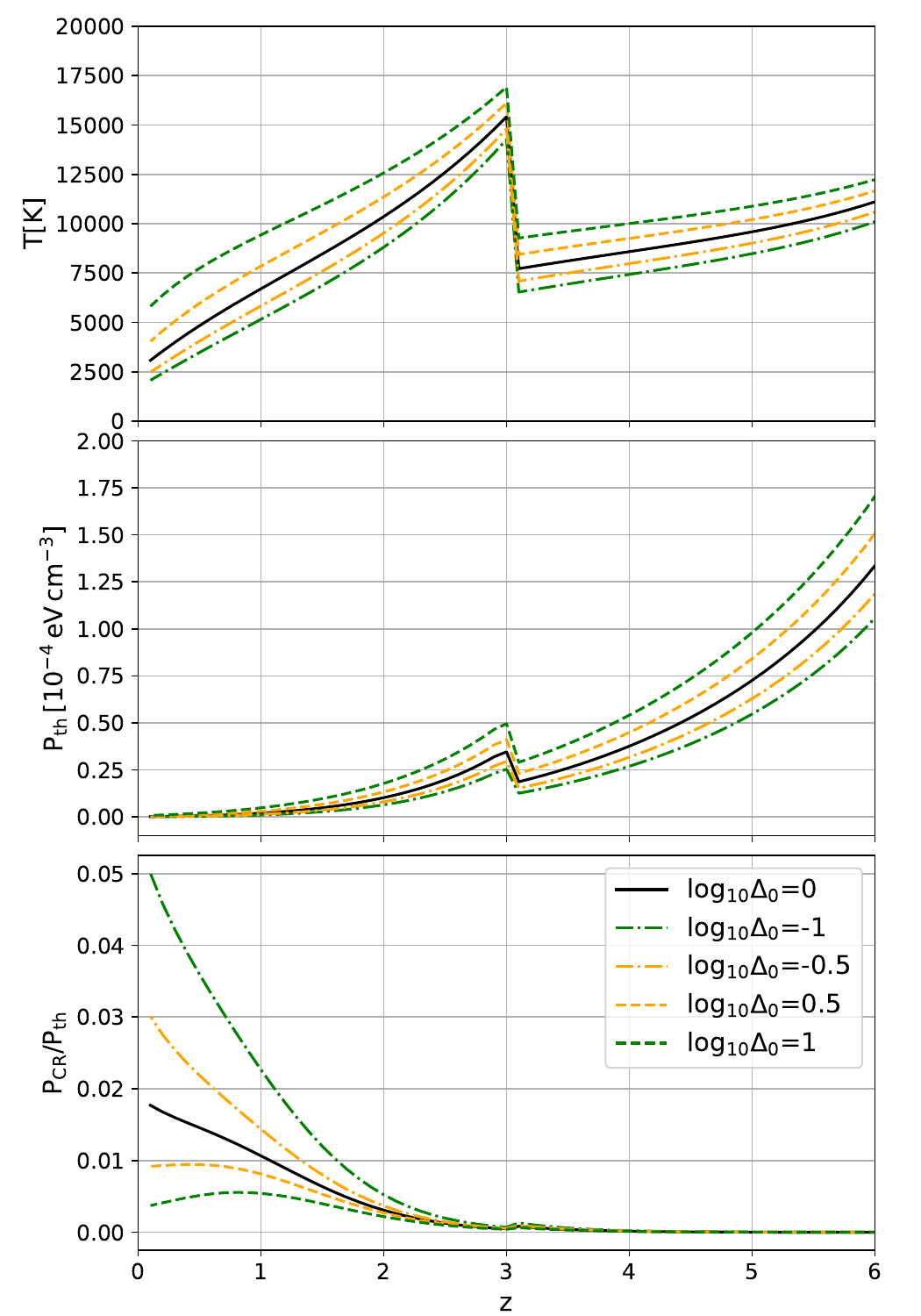}
\caption{The temperature profile (top), IGM thermal pressure (middle), and the pressure comparison (bottom) with redshift for different overdensity as labeled. The temperature experiences a sudden increase at $z$ = 3 (observed towards $z$ = 0) due to the instantaneously ionization of He\,{\sc ii}. Which is also responsible for the rigid bump for the other two panels. In general, the temperature drops toward lower redshift, and higher overdensity corresponds to higher temperature profile (a ``normal'' temperature-density relation). The fractional contribution of cosmic ray pressure to the general IGM pressure progressively increases toward lower redshifts (again, with a small perturbation at $z=3$ due to helium reionization), and $P_{\rm CR}/P_{\rm th}$ is also largest in the low density regions (where the energy range over which cosmic ray electrons can survive without collisional cooling is largest).
\label{fig:TP}}
\end{figure}

To illustrate the additional pressure introduced by the MeV electron cosmic ray in the IGM, we model the thermal and pressure evolution of IGM. The temperature evolution of IGM gas is described by the following equation \citep{1997MNRAS.292...27H}:
\begin{equation}
    \frac{dT}{dt} = -2HT + \frac{2T}{3\Delta}\frac{d\Delta}{dt}-\frac{T}{n_{\rm tot}}\frac{dn_{\rm tot}}{dt} + \frac{2}{3k_{\rm B}n_{\rm tot}}\frac{dQ}{dt}
\end{equation}
where $n_{\rm tot}$ is the total number density of free baryonic particles, $k_{\rm B}$ is the Boltzmann constant. The term $dQ/dt$ encodes the heating and cooling processes in the IGM and in this work we follow the modeling details in \citet{2016MNRAS.460.1885U}. 
The fiducial reionization scenario we choose is that H\,{\sc i} reionization happens instantly at $z=8$, heating the gas to $T_0=2\times 10^4\, K$, and that He\,{\sc ii} reionization happens at $z=3$ also instantaneously heats IGM through photoheating. We then calculate the thermal pressure of IGM using the ideal gas law, $P_{\rm th}(z) = n_{\rm tot}(z)k_{\rm B}T(z)$.

Figure \ref{fig:TP} illustrates the temperature profile, fiducial IGM pressure, and the evolution of pressure ratio of MeV electron CR to IGM with respect to redshift for different overdensity settings. The temperature experiences a sudden increase at $z$ = 3 (observed towards $z$ = 0) due to the instantaneously ionization of He\,{\sc ii}. Generally, the temperature drops from $z$ = 15 to $z$ = 0, and higher overdensity corresponds to higher temperature profile. The general IGM pressure exhibits a decrease from $z$ = 15 to $z$ = 0, with higher overdensity resulting in higher pressure. Moreover, the contribution of cosmic ray pressure to the general IGM pressure progressively increases from $z$ = 15 to $z$ = 0, while the percentage decreases as the overdensity increases. At mean density, the ratio of cosmic ray pressure to IGM pressure is 0.3\% at $z$ = 2, 1\% at $z$ = 1, and 1.8\% at $z$ = 0.1.

\subsection{The overall electron spectrum}
\label{ss:espec}

We may also use our results to compute the overall electron spectrum in the IGM, ranging from the thermalized electrons at $\sim$eV energies to the Compton-induced cosmic rays that peak at $\sim$MeV energies. The transition range between the thermal and cosmic ray contributions contains a rather small amount of energy and hence is not directly relevant to the energetics or pressure of the IGM. However, we will see in Paper II that it can dominate the damping of plasma waves at some wave numbers, so it is important to fill in the transition regime to get a complete description of IGM electrons.

To do this, it is necessary to both run our calculation down to lower energies, and include one more contribution: the transient electrons generated by photoionization, which may be generated at energies that extend far beyond the tail of the Maxwell-Boltzmann distribution and then cool rapidly. After \ion{He}3 reionization, the major processes of photoionization equilibrium in the IGM are
\begin{equation}
\left\{ \begin{array}l
{\rm H}^+ + e^- \rightarrow {\rm H(1s)} + \gamma\, (\,+\gamma ...) \\
{\rm H(1s)} + \gamma_{\rm UVB} \rightarrow {\rm H}^+ + e^-
\end{array}\right.
~~~{\rm and}~~~
\left\{ \begin{array}l
{\rm He}^{2+} + e^- \rightarrow {\rm He^+(1s)} + \gamma\, (\,+\gamma ...) \\
{\rm He^+(1s)} + \gamma_{\rm UVB} \rightarrow {\rm He}^{2+} + e^-
\end{array}\right..
\end{equation}
where the top reaction is recombination (there are additional photons emitted if the initial recombination is to an excited state). The overall rates of these reactions in units of events cm$^{-3}$ s$^{-1}$ are $\alpha_{{\rm A},X}(T_{\rm IGM}) n_e n_X$, where $X$ is one of H or He and $\alpha_{\rm A}$ is the Case A recombination coefficient for the dominant fully ionized species (we use the fits of \citet{1991A&A...251..680P}). The photoionization immediately follows: the rate of generation of electrons with energy $>E$ (in units of cm$^{-3}$ s$^{-1}$) is then $\alpha_{{\rm A},X} n_e n_X$ times the fraction of photoionizations of species $X$ that lead to an energy $>E$:
\begin{equation}
\dot N_{\rm pi}(>E) = \sum_{X\in\rm\{H,He\}} \alpha_{{\rm A},X} n_e n_X \frac{\int_{(I_X+E)/h}^\infty \frac{4\pi}{h\nu} \sigma_{{\rm pi},X}(\nu) J_{\nu}\, d\nu}{\int_{I_X/h}^\infty \frac{4\pi}{h\nu} \sigma_{{\rm pi},X}(\nu)  J_{\nu}\, d\nu }.
\end{equation}
Here $J_\nu$ is the EBL spectrum and $\sigma_{{\rm pi},X}$ is the photoionization cross section of species $X$ (obtainable from the standard techniques for hydrogenic species, e.g., \citealt{1957qmot.book.....B}).

The photoionization-induced electrons rapidly dissipate their energy through 2-body collisions. We estimate their energy loss through the usual non-relativistic 2-body plasma formula
\begin{equation}
-\left.\frac{dE}{dt}\right|_{\rm loss} = \frac{4\pi e^4 n_e}{m_ev} \ln \Lambda,
\end{equation}
where $\ln \Lambda$ is the Coulomb logarithm (here we take $\Lambda = b_{\rm max}/b_{\rm min}$, where $b_{\rm min} = e^2/E$ is set by large deflections and $b_{\rm max} = k_{\rm D}^{-1}$ is at the Debye scale). The total photoionization-induced spectrum at any given time is
\begin{equation}
N_{\rm pi}(E) = \frac{\dot N_{\rm pi}(>E)}{-(dE/dt)|_{\rm loss}}.
\label{eq:NPI}
\end{equation}
Once the electrons cool to the thermal energy, this formula ceases to be accurate: then the electrons start to gain energy due to kicks from neighboring electrons, and eventually merge into a Maxwell-Boltzmann distribution in accordance with the fluctuation-dissipation theorem. But by the time this happens, the transient photoionization contribution is negligible compared to the population of thermal electrons.

\begin{figure}
\includegraphics[width=\textwidth]{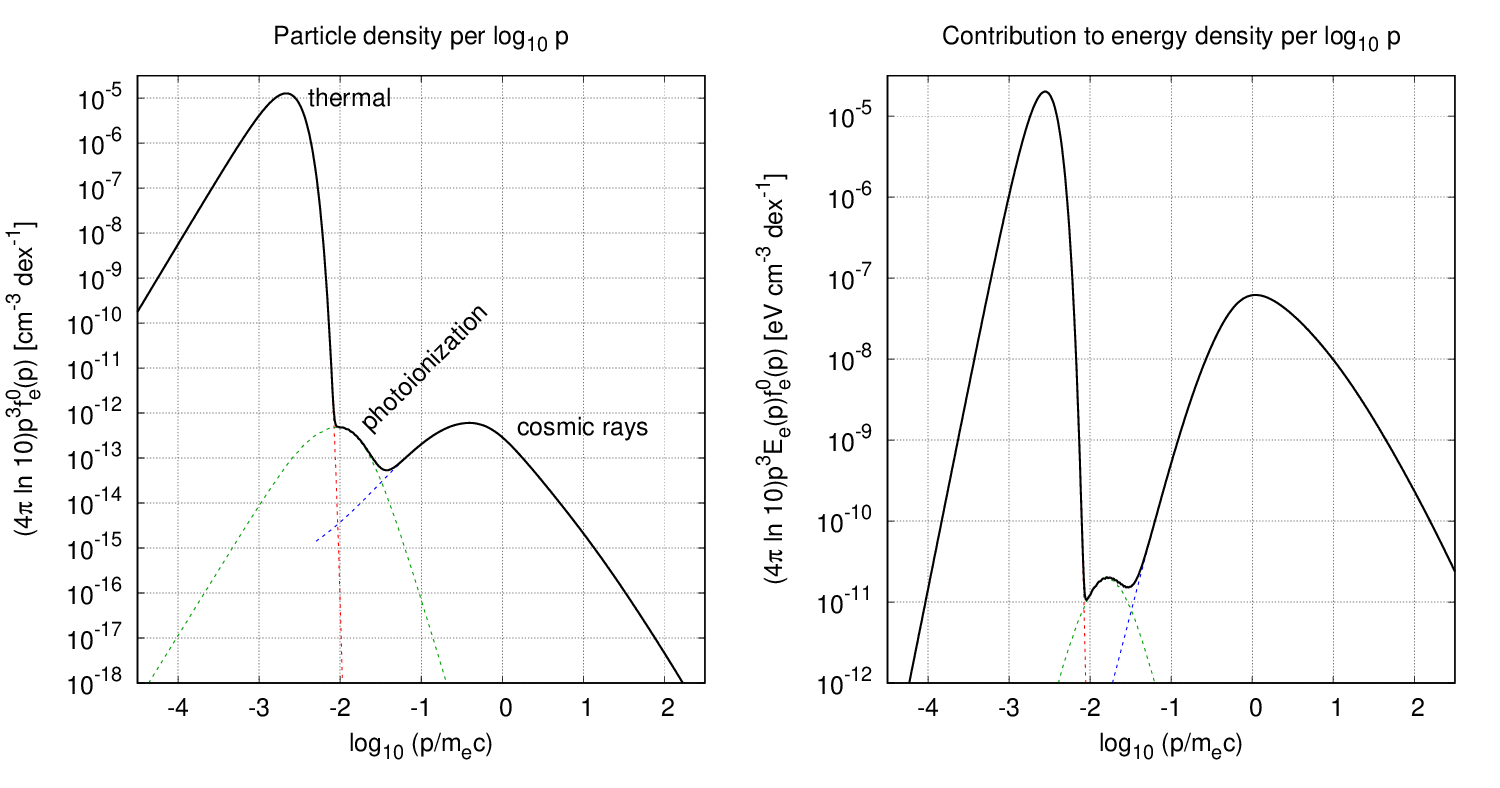}
\caption{\label{fig:phasespace}The phase space density of electrons at $z=2$ in the mean-density IGM. We show the thermal (Maxwell-Boltzmann), photoionization (Eq.~\ref{eq:NPI}), and cosmic ray components. The left panel is stretched to show particles per logarithmic range in momentum, while the right panel is weighted by energy.}
\end{figure}

In Fig.~\ref{fig:phasespace}, we plot the results in the form most useful for plasma theory, which is the phase space density
\begin{equation}
f_e(p) = \frac1{4\pi p \sqrt{m_e^{\changetextmath 2}+(p/c)^2}}\, N\left( E = \sqrt{(pc)^2+(m_ec^2)^2}-m_ec^2 \right).
\label{eq:fep}
\end{equation}

\subsection{Effect on plasma waves and ultrarelativistic beam instabilities}
\label{ss:beam}

The plasma instabilities that are proposed to be relevant for TeV pair beams are associated with the growth of electrostatic plasma waves. An unmagnetized can undergo electrostatic oscillations at a frequency of
\begin{equation}
\omega_{\rm p} = c \sqrt{4\pi r_0n_e}
= 139\chi^{1/2} z_2^{3/2} \Delta^{1/2}\,{\rm s}^{-1},
\end{equation}
where $z_2\equiv (1+z)/3$ and $\chi$ is the ionization fraction (normalized to 1 for fully ionized H$^+$ and He$^{2+}$). In the cold limit, the dispersion relation is truly flat, $\omega = \omega_{\rm p}$, the phase velocity is $v_{\rm ph} = \omega_{\rm p}/k$, and the group velocity is zero. 
The deviation of the plasma from being perfectly cold has two major effects. First, there is a correction to the real part of the frequency, $\Re\omega(k) - \omega_{\rm p}$, although this is small for the waves with $v_{\rm ph}/c$ of order unity that are relevant here. Secondly, there is a correction to the imaginary part $\Im\omega(k)$, which results in plasma waves growing ($\Im\omega>0$) or decaying ($\Im\omega<0$). We start from the form given by \citet{1995PhRvL..74.2232B}, which is valid for small numbers of non-thermal particles, including in the relativistic regime, with some algebraic rearrangement:
\begin{equation}
\Im \omega(k) = \frac{\pi\omega_{\rm p}}{2n_e} \left( \frac{m_e\omega_{\rm p}}{k} \right)^2
\int_{{\mathbb R}^3} d^3{\bf p}\,
\,\delta \left(m_e\omega_{\rm p} - \frac{{\bf k}\cdot{\bf p}}\gamma\right)
\,{\bf k}\cdot\nabla_{\bf p}f_e({\bf p}),
\label{eq:Im-omega}
\end{equation}
where ${\bf p}$ is the electron momentum and $f_e({\bf p})$ is the phase space density. This correction is often negative (Landau damping), but can be positive if a beam is passing through the plasma (see \citealt{2010PhPl...17l0501B} for a review). For a narrow ultrarelativistic relativistic beam, $p\approx m_ec\gamma$, and the resonance condition in the $\delta$-function implies that modes with
\begin{equation}
k \approx \frac{\omega_{\rm p}}c \sec\theta
\label{eq:k-res}
\end{equation}
will be most strongly affected by the beam, where $\theta$ is the angle between ${\bf k}$ and the beam direction (see \citealt{2010PhRvE..81c6402B, 2016ApJ...833..118C} for some fully worked examples). \changetext{The ``width'' $\Delta k$ in which significant growth occurs depends on the beam opening angle (i.e., the spread in propagation directions of the electrons and positrons); this is determined by the pair production cross section \citep{2012ApJ...758..101S} and broadening by any pre-existing intergalactic magnetic field (see \S\ref{ss:preigmf}) or by plasma instabilities themselves (see, e.g., \citealt{2024arXiv240203127A} for a detailed recent study).}

If we take the isotropic population of $\sim\,$MeV cosmic rays to be isotropic, we may solve the problem in spherical polar coordinates, with $\tilde\mu$ as the cosine of the angle between ${\bf p}$ and ${\bf k}$: $d^3{\bf p} \rightarrow 2\pi p^2\,dp\,d\tilde\mu$. Then the $\tilde\mu$ integral is collapsed by the $\delta$ function at $\tilde\mu = \omega_{\rm p}/(kv)$ and we have a contribution to the growth rate:
\begin{eqnarray}
\Im \omega &=& 
\frac{\pi^2\omega_{\rm p}}{n_e} \left( \frac{m_e\omega_{\rm p}}{k} \right)^2 \int_0^\infty p^2\,dp \int_{-1}^1 d\tilde\mu\, \delta(m_e(\omega_{\rm p} - kv\tilde\mu))\,k\tilde\mu \frac{df_{e}(p)}{dp}
= \pi^2\omega_{\rm p} \left( \frac{m_e\omega_{\rm p}}{k} \right)^3
\int_{p_{\rm min}}^\infty dp\, 
\left[ 1 + \frac{p^2}{(m_ec)^2} \right] \frac{df_{e}(p)}{dp}
\nonumber \\
&=& -\frac{\pi^2\omega_{\rm p}}{n_e} \left( \frac{m_e\omega_{\rm p}}{k} \right)^3 \left\{ \left[ 1 + \frac{p_{\rm min}^2}{(m_ec)^2} \right]f_{e}(p_{\rm min})
+ \frac{2}{(m_ec)^2}\int_{p_{\rm min}}^\infty pf_{e}(p)\,dp
\right\},
\label{eq:temp-imomega}
\end{eqnarray}
where $v=p/\sqrt{p^2+(m_ec)^2}$ is the particle velocity; the minimum momentum of particles that can resonate with the wave is $p_{\rm min} = m_ev_{\rm min}/\sqrt{1-v_{\rm min}^2/c^2}$ and $v_{\rm min} = \omega_{\rm p}/k$; and in the last line we have integrated by parts. The resonance condition of course implies that $\Im\omega=0$ if $k<\omega_{\rm p}/c$. If we substitute the phase space density for $N(E)$ given by Eq.~(\ref{eq:fep}), and putting in the appropriate restriction on the range of $k$ using the Heaviside $\Theta$-function, Eq.~(\ref{eq:temp-imomega}) simplifies to
\begin{equation}
\Im \omega
= -\frac\pi 4 \omega_{\rm p} \left( \frac{\omega_{\rm p}}{ck} \right)^3 \frac{m_ec^2}{n_{e,\rm th}} \left\{ \frac{E_{\rm min}+m_ec^2}{\sqrt{E_{\rm min}(E_{\rm min}+2m_ec^2)}} N(E_{\rm min})
+ 2\int_{E_{\rm min}}^\infty \frac{N(E)\,dE}{\sqrt{E(E+2m_ec^2)}}
\right\} \Theta\left( \frac{ck}{\omega_{\rm p}}- 1\right)
\label{eq:ImCR}
\end{equation}
with $E_{\rm min} = m_ec^2 [(1-\omega_{\rm p}^2/c^2k^2)^{-1/2}-1]$ being the kinetic energy of an electron whose velocity matches the phase velocity of the wave.
The $-$ sign indicates that the plasma oscillations are damped. This reduces to the usual non-relativistic expression for Landau damping \citep{1960JNuE....1..171J} when $E\ll m_ec^2$; in this case only the first term in braces is significant.

\begin{figure}
\centering{\includegraphics[width=0.47\textwidth]{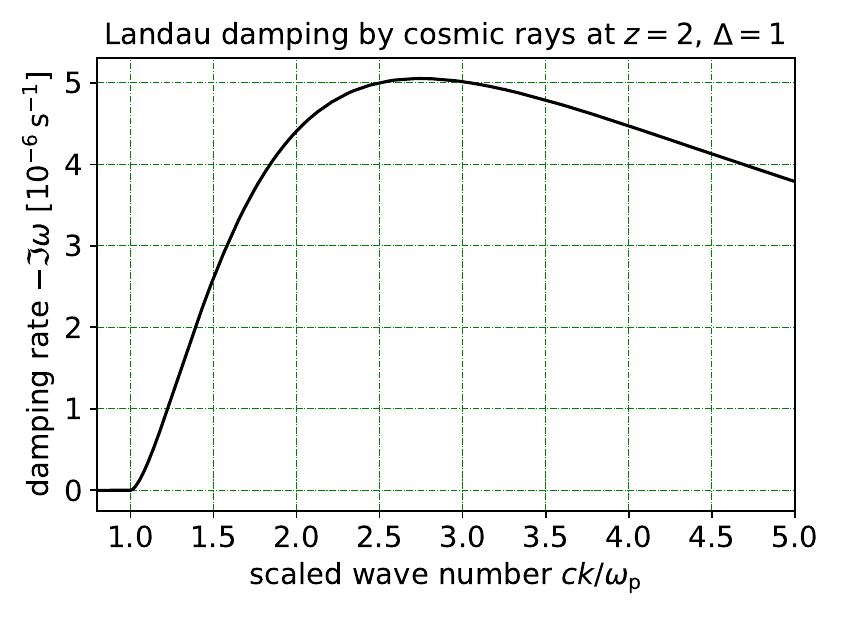}
\includegraphics[width=0.5\textwidth]{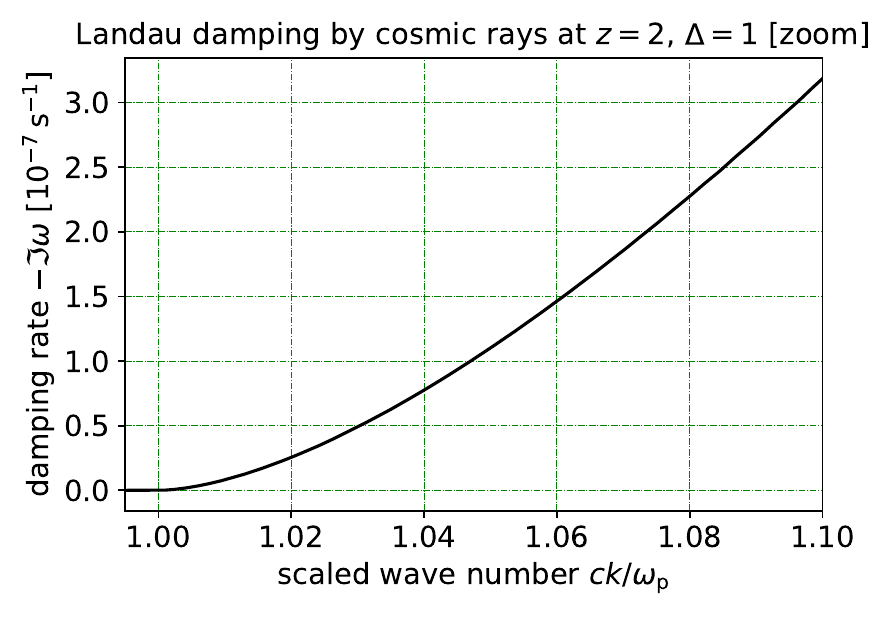}}
\caption{\label{fig:Landau}The linear Landau damping rate for plasma oscillations at $z=2$ and $\Delta=1$. The wavenumber is scaled by $\omega_{\rm p}/c = 4.6\times 10^{-9}\,$cm$^{-1}$. Note that there is no linear Landau damping for phase velocities exceeding $c$ or $k<\omega_{\rm p}/c$.}
\end{figure}

The linear Landau damping rate caused by the $\sim$MeV cosmic rays is plotted in Fig.~\ref{fig:Landau} for mean density at $z=2$, and for $1.2<ck/\omega_{\rm p}\lesssim 5$ it is a few$\times 10^{-6}\,$s$^{-1}$. For comparison, the instability growth rate $\Gamma_{\rm M,k}$ predicted by \citet{2012ApJ...752...22B} is $\approx 2.7\times 10^{-11} E_{\rm TeV}^{3/2}\Delta^{-1/4}$\,s$^{-1}$ for their fiducial blazar ($EL_E=10^{45}$ erg s$^{-1}$) at $z=2$ and at a distance of 1 pair production path length from the blazar; and the electron-ion collision rate is
\begin{equation}
\tau^{-1}_{ei} =\sqrt{\frac{32\pi}9}\, Z n_e r_0^2c^4 \left(\frac{k_{\rm B}T}{m_e}\right)^{-3/2} \ln\Lambda = 7.2\times 10^{-10} z_2^3 \Delta T_4^{-3/2} \left(\frac{\ln\Lambda}{29.2}\right)~{\rm s}^{-1}
\label{eq:tauei}
\end{equation}
\citep[e.g.][Eq.~2.5e]{1965RvPP....1..205B}, where $Z$ is the weighted mean charge of the ions (1.14 after helium reionization), $T_4$ is the IGM temperature in units of $10^4$ K, and $\ln\Lambda$ is the Coulomb logarithm. It is readily seen that for $k\sim $few$\times \omega_{\rm p}/c$, the linear Landau damping operates at least thousands of times faster than either the pair beam instability or collisional effects. We thus conclude that in this range of wavenumbers, the oblique instability is simply shut off. The linear Landau damping rate is so enormous that even order-of-magnitude revisions in the MeV gamma ray background or the redshift dependence of its emissivity would not change the conclusion.

However, the resonance condition in Eq.~(\ref{eq:k-res}) does allow a way out: at small angles, $ck/\omega_{\rm p} = 1 + \frac12\theta^2 + ...$, the damping rate declines dramatically as $ck/\omega_{\rm p}\rightarrow 1$ (see the right-hand panel of Fig.~\ref{fig:Landau}). The essential reason is that plasma modes with small $\theta$ can resonate with the ultra-relativistic $\sim$TeV leptons in the pair beam ($v_{\rm ph} \approx c \cos\theta$), but can outrun the quasi-relativistic cosmic ray electrons, $v_{\rm ph}>c\beta$. \changetext{The outcome of growth of these linear modes is a matter of debate, with proposals ranging from thermalization of a large portion of the beam energy to saturation of the instability via non-linear damping of plasma waves and/or broadening of the beam and consequent suppression of the growth rate. A wide range of effects on the observed gamma ray spectra can occur, even for the same modeling assumptions but different blazar properties and IGM conditions \citep{2019MNRAS.489.3836A}. We plan to investigate the growth of the very small-$\theta$ modes, and whether they can lead to dramatic angular broadening or partial thermalization of the beam energy, in future work.}

\subsection{Order-of-magnitude estimation of magnetic field generation}
\label{app: B_OOM}

The cosmic rays considered in this paper should be considered as a potential source of magnetic fields. There are two classes of sources for magnetic fields in an initially unmagnetized plasma. The ``battery'' mechanisms deterministically generate a large-scale ${\boldsymbol B}$-field from inhomogeneities in a multi-component plasma. In ``instability'' mechanisms, the free energy in an anisotropic particle distribution can drive an instability on small scales (seeded by Poisson fluctuations if nothing else is available), resulting in a stochastic ${\boldsymbol B}$-field that grows exponentially until non-linear effects cause saturation. The classic examples of these mechanisms are the Biermann battery in an electron-ion plasma with misaligned temperature and density gradients \citep{1950ZNatA...5...65B}, and the Weibel instability in a plasma with quadrupolar anisotropy of the electron velocity distribution \citep{1959PhRvL...2...83W}.

We consider the magnetic field in the IGM generated by the electron cosmic rays via the return current battery mechanism \citep{2020ApJ...896L..12O}. The mechanism works slightly differently in the case that the thermal component is collisional on the cosmic ray transport timescale, as in the case of the IGM \citep{2011ApJ...729...73M}.\footnote{The collision time noted above is $\sim 10^9\,$s, whereas even at the speed of light a cosmic ray can traverse the IGM pressure-smoothing scale of $\sim 100$ kpc in $\sim 10^{12}$\,s.} The key is that there will be a return current $-{\boldsymbol J}_{\rm CR}$ in the thermal medium, which implies an electric field ${\boldsymbol E} = \sigma_{\rm cond}^{-1}{\boldsymbol J}$ by Ohm's law. Faraday's law then implies that the curl component of this electric field is associated with a time-dependent magnetic field:
\begin{equation}
    \frac{\partial {\boldsymbol B}}{\partial t} = c\,{\boldsymbol\nabla}\times (\sigma^{-1}_{\rm cond}{\boldsymbol J}_{\rm CR}),
\end{equation}
where $\sigma_{\rm cond}$ is the conductivity of IGM and ${\boldsymbol J}_{\rm CR}=-e\,n_{\rm CR}{\boldsymbol v}_{\rm CR,bulk}$ is the electron cosmic ray current density. 

\begin{figure}
\centering{\includegraphics[width=0.6\textwidth]{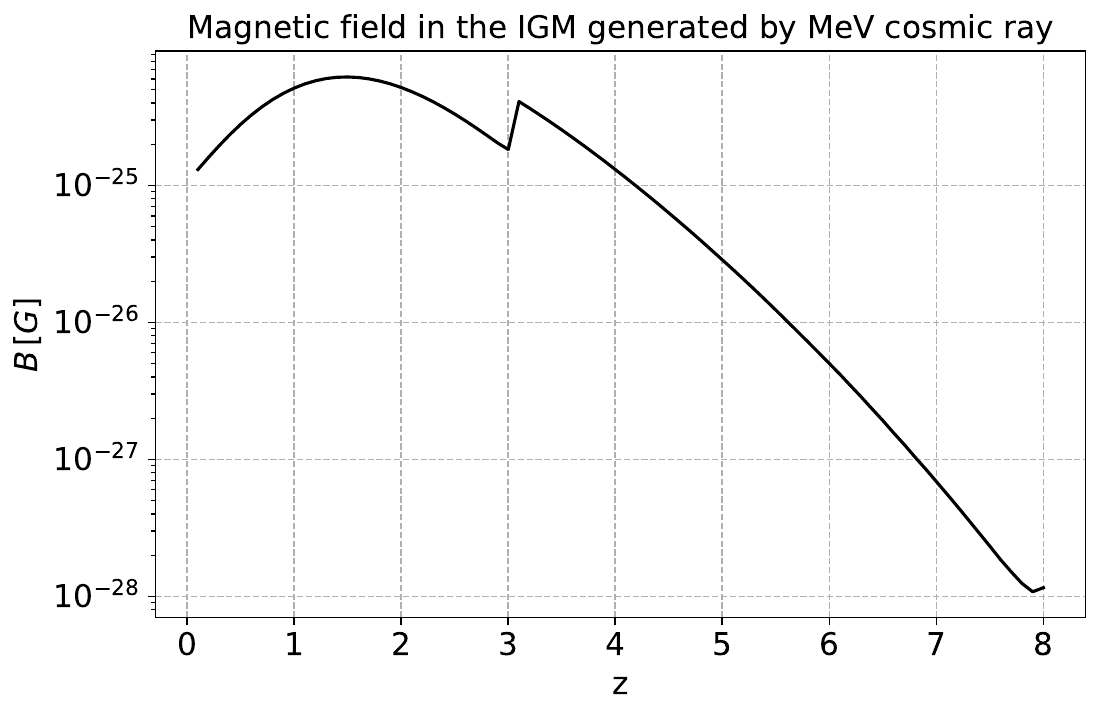}}
\caption{\label{fig:B_max} An order-of-magnitude estimate of the magnetic fields in the IGM generated by electron cosmic rays via the return current battery mechanism.}
\end{figure}

Assuming the cosmic ray can diffuse in the Universe, we have the approximation $(v_{\rm bulk}\Delta t)/L\sim \mathcal{O}(1)$, where $L$ is the characteristic scale that matter density varies; this would be of order the filtering or the Jeans scale in the IGM. We also expect that when density perturbations are of order unity or greater, there is (at order of magnitude level) no special preference for the rotational vs.\ irrotational components of $\sigma^{-1}_{\rm cond}{\boldsymbol J}_{\rm CR}$. Therefore, we have
\begin{equation}\label{eq:B_max}
B\sim\frac{c\,e\, n_{\rm CR}}{\sigma_{\rm cond}}.
\end{equation}
If the plasma is initially unmagnetized, we take the Spitzer conductivity \citep{1953PhRv...89..977S}:
\begin{equation}
    \sigma_{\rm cond} = \frac{3}{4\sqrt{2\pi}}\frac{(k_{\rm B}T)^{3/2}}{Z e^2m_{\rm e}^{1/2}\ln{\Lambda}}\frac{1}{F(Z)},
\end{equation}
where $k_{\rm B}$ is Boltzmann constant, $\ln{\Lambda}\sim14$ is the Coulomb logarithm, for electron-hydrogen collision $Z=1$ and $F(Z)\sim 1/1.96$.
Plugging into Eq.~(\ref{eq:B_max}) the profiles of IGM temperature, number densities of cosmic rays we have in Sec.~\ref{sec:res}, we estimate the magnetic field as a function of redshift, shown in Fig.~\ref{fig:B_max}. The results show that the order of magnitude of this magnetic field is $\sim 10^{-25}$\,G at $z\lesssim 4$. This is too small to have an impact on any known observable --- indeed, the implied cyclotron frequency is approximately the Hubble rate. It is also small compared to the proposed field generation mechanisms from cosmic rays from early galaxies \citep{2011ApJ...729...73M}, although the latter requires the cosmic rays to be able to freestream into the IGM without being confined by instabilities. So we do not expect the return current battery driven by the Compton-sourced electron cosmic rays to be significant.

An alternative is to consider the quadrupole or \citet{1959PhRvL...2...83W}-type instability mechanisms, which operate on an initially unmagnetized plasma.
In the collisionless case, there is an instability for any non-zero quadrupole anisotropy in the particle momentum distribution, even when relativistic effects are considered \citep{1969JMP....10...13L}. We defer a full analysis for future work and focus here only on the orders of magnitude. As such, we will use the non-relativistic formulation, as described in \citet{1962JFM....14..321K, 1964JFM....19..210K}. If there is a cosmic ray component with number density $n_{e,\rm CR}$ and fractional quadrupole anisotropy $A_{\rm CR}$, and a thermal electron component with density $n_{e,\rm th}\gg n_{e,\rm CR}$, then the characteristic wavenumber $k_\star$ of the Weibel instability and its growth rate $\Gamma_\star$ are of order\footnote{For transverse waves along a principal axis of the velocity ellipsoid, the dispersion relation reduces to $k^2/k_0^2 + I_{ww}(a) + 1 = 0$, where $a$ is the phase velocity in the notation of \citet{1962JFM....14..321K} and $k_0=\omega_{\rm p}/c$. If one follows the Taylor expansion in $a$, $I_{ww}(a) = I_{ww}(0) + I'_{ww}(0)a + ...$, then the cosmic rays determine $1+I_{ww}(0)$ since the latter vanishes for any isotropic distribution; but the thermal electrons dominate $I'_{ww}(0)$.}
\begin{equation}
k_\star \sim \left( A_{\rm CR} \frac{n_{e,\rm CR}}{n_{e,\rm th}} \right)^{1/2}\frac{\omega_{\rm p}}c
~~~{\rm and}~~~
\Gamma_\star \sim \left( A_{\rm CR} \frac{n_{e,\rm CR}}{n_{e,\rm th}} \right)^{3/2}\frac{v_{e,\rm th}}c \omega_{\rm p},
\end{equation}
where $v_{e,\rm th}$ is the electron thermal velocity. At $z\sim 2$, we estimate the growth rate to be $\sim 10^{-15}A_{\rm CR}^{3/2}\,$s$^{-1}$. It is not clear what quadrupole anisotropy to expect for the cosmic rays, and it presumably depends on where one is between the diffusion and free-streaming limits. However, even a few percent-level anisotropy is not sufficient to drive the Weibel instability in less than a Hubble time, and regardless of $A_{\rm CR}$ the timescales are long enough for collisions to be relevant. We therefore expect that the Weibel instability driven by Compton-induced cosmic ray electrons is not likely to be operative.

\subsection{Pre-existing intergalactic magnetic fields}
\label{ss:preigmf}

\changetext{Both the aforementioned growth rate of the beam-plasma instability \citep{2012ApJ...752...22B} and the Landau damping rate in \S\ref{ss:beam} assume a negligible pre-existing intergalactic magnetic field (IGMF). This may well be the case in ``undisturbed'' regions of the IGM that have not yet passed through a structure formation shock. However, if an IGMF is present, it could broaden or even isotropize the angular dispersion of the TeV pair beams, which both modifies the plasma instability growth rate and changes the morphology of the cascade (inverse Compton) gamma rays from pair beams \citep{1994ApJ...423L...5A}. The latter effect makes TeV blazars a potential probe of IGMFs (see, e.g., \citealt{2021Univ....7..223A} for a review). In this section, we briefly discuss these issues, and also check whether the IGMF could modify the Landau damping rate from the MeV cosmic rays.}

\changetext{Electrons and positrons in the pair beam produced by a blazar are directed forward, with a typical initial opening angle of $\sim 1/\gamma$ in accordance with relativistic kinematics (see \citealt{2012ApJ...758..101S} for some detailed distributions), and with an inverse Compton cooling length of
\begin{equation}
\ell_{\rm IC} = \frac{E}{-\dot E} = \frac{3m_ec^2}{4a_{\rm rad}T_{\rm CMB}^4\sigma_{\rm T}\gamma}
= 1.4\times 10^{22}z_2^{-4} E_{\rm TeV}^{-1} \,{\rm cm}
\end{equation}
(this is $c/\Gamma_{\rm IC}$ in the notation of \citealt{2012ApJ...752...22B}).
If there is a magnetic field ${\boldsymbol B}$ at an angle $\zeta$ to the pair beam, that is coherent on the scale of the cooling length $\ell_{\rm cool}$ (which is $\ell_{\rm IC}$ if inverse Compton cooling is the main process), then the leptons can be deflected through an angle
\begin{equation}
\Delta\theta_B \sim \frac{eB\ell_{\rm cool}}{E} \sin\zeta
= 4.2\times 10^{-3} z_2^{-4} E_{\rm TeV}^{-2} B_{\rm fG} \frac{\ell_{\rm cool}}{\ell_{\rm IC}}\sin\zeta\,{\rm radian}
\label{eq:dtb}
\end{equation}
before they cool. Even a uniform field leads to an angular broadening of the beam because some charged particles were ``recently'' produced, whereas others have traversed a distance of $\sim \ell_{\rm cool}$; but the broadening is only on one of the two axes (``East-West'' with respect to the field direction). A magnetic field with a coherence length shorter than $\ell_{\rm cool}$ would lead to smaller total deflection than in Eq.~(\ref{eq:dtb}), but the broadening would be on both axes. A magnetic field below $1.2\times 10^{-19}E_{\rm TeV}z_2^{-4}\,$G leads to $\gamma\Delta\theta_B<1$ even if $\ell_{\rm cool}=\ell_{\rm IC}$; at field strengths higher than this, the IGMF should be considered as a potential broadening mechanism.}

\changetext{If the IGMF is sufficiently weak, and the plasma instabilities are inactive, then the cascade photons are beamed forward and would appear coincident with the source; at somewhat higher field strengths they might appear in a ``halo'' (short coherence length) or ``bowtie'' (long coherence length; \citealt{2015JCAP...09..065L, 2017ApJ...850..157T}). Limits on these effects in low-redshift TeV blazars have been used to constrain the IGMF strength and coherence length \citep[e.g.][]{2010Sci...328...73N, 2010A&A...524A..77A, 2012ApJ...747L..14V, 2014A&A...562A.145H, 2017ApJ...835..288A, 2018ApJS..237...32A, 2020ApJ...892..123T, 2023ApJ...950L..16A}.}

\changetext{The broadening of the beam by an IGMF would reduce the rate of growth of plasma instabilities, since the maximum growth rate for oblique instability of an ultra-relativistic beam of density $n_{\rm beam}$ is
\begin{equation}
\Gamma_{\rm growth} \sim 0.4 \omega_{\rm p} \frac{n_{\rm beam}}{n_{e,\rm th}} \frac1{\sigma^2_\theta},
\end{equation}
where $\sigma^2_\theta = k_{\rm B}T_{\rm beam}/(m_ec^2\gamma_{\rm beam})$ is the squared width of the beam and the factor of 0.4 is appropriate for a Maxwell-J\"uttner distribution function \citep[e.g.][]{2016ApJ...833..118C}. \citet{2012ApJ...752...22B} recognized that beam angular broadening by an IGMF was a potential mechanism to shut off the plasma instabilities.
In order to slow the growth of all relevant modes, the beam must be angularly broadened on both axis, as happens naturally if there is an random component on scales $\sim\ell_{\rm IC}$ \citep{2022ApJ...929...67A}.}

\changetext{In addition to broadening the beam and slowing the growth of plasma instabilities, magnetic fields may modify the Landau damping of plasma wave modes --- indeed, the entire concept of a ``resonant'' electron changes when a background magnetic field is introduced, because the unperturbed trajectory of an electron is now a helix instead of a straight line. Therefore, one might wonder whether there is a range of field strengths that could disrupt the linear Landau damping by MeV cosmic rays, but leave the instability of the TeV beam intact. The Landau damping in the presence of a magnetic field is discussed in Appendix~\ref{app:rlm}, and can be evaluated using Eq.~(\ref{eq:growth-mag}).}

\begin{figure}
\includegraphics[width=\textwidth]{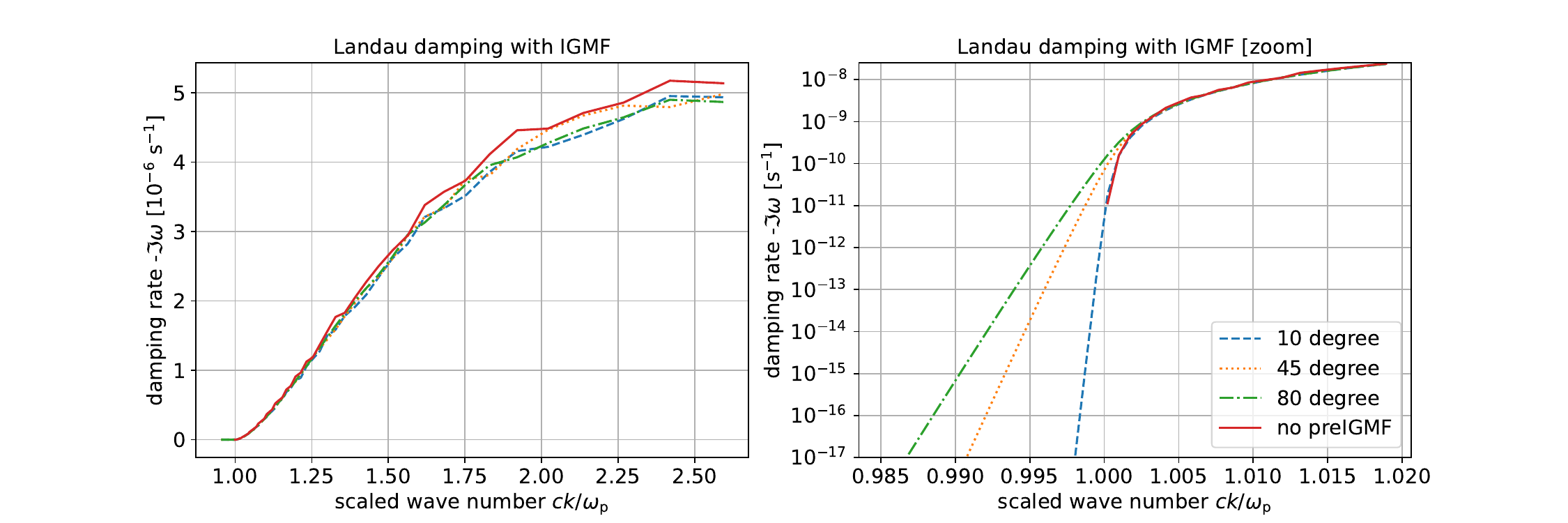}
\caption{\label{fig:mag-damp}\changetext{The linear Landau damping rates at $z=2$ and mean density with and without a background magnetic field is included. The magnetic field strength used was the equipartition value, 20.8 nG. Note that there is only a minor change in the results at $k\gtrsim 1.002\omega_{\rm p}/c$; at smaller wavenumbers, the case with the magnetic field exhibits an exponentially declining ``tail.'' Curves are presented for an angle between the wavenumber and magnetic field direction of $\Theta = 10$, 45, and 80 degrees.}}
\end{figure}

\changetext{The damping rates at $z=2$ and mean density are shown in Fig.~\ref{fig:mag-damp} for two cases: no background magnetic field (as assumed in \S\ref{ss:beam}) and an equipartition field, with magnetic energy density equal to the thermal energy density (in this case: $B=20.8\,$nG). The equipartition field is marginally consistent with current upper limits from Faraday rotation \citep[e.g.][]{2022MNRAS.515..256P}, depending on other parameters such as coherence length. Even at this field strength, the Landau damping at $k\gtrsim 1.002\omega_{\rm p}/c$ is only weakly affected. At $k<\omega_{\rm p}/c$, there is no linear Landau damping without a magnetic field, because there are no particles that resonate with these waves (the phase velocity is faster than the speed of light). The presence of a magnetic field allows charged particles traveling slower than the phase velocity $\omega/k$ to absorb wave energy: this is similar to what happens in synchrotron absorption, except that in our case the wave is a mostly-longitudinal plasma oscillation instead of an electromagnetic wave. This has a dependence on angle $\Theta$ of the wave vector ${\boldsymbol k}$ relative to the magnetic field ${\boldsymbol B}$, with more damping in nearly perpendicular directions ($\Theta=80$ degrees) than nearly parallel directions ($\Theta=10$ degrees). But in any case the damping rates rapidly become less than the collisional damping rate, Eq.~(\ref{eq:tauei}). We conclude that the Landau damping mechanism is robust against even large magnetic fields approaching the equipartition value (which, in accordance with Eq.~\ref{eq:dtb} and $\sim 2\times 10^7$ fG, would completely disrupt the TeV pair beam unless it happened to be pointed almost exactly along the ${\boldsymbol B}$-field direction).}

\section{Discussion}
\label{sec:fin}

We have shown in this paper that cosmic ray electrons are ubiquitous in the IGM. They contribute about 0.3\% to thermal pressure in the IGM at $z$ = 2 at mean density, rising to 1.0\% at $z$ = 1 and 1.8\% at $z$ = 0.1.
Our study also reveals that cosmic ray electrons induce linear Landau damping of plasma oscillations at a rate significantly faster than previously predicted instability and collision rates. This damping effectively suppresses the oblique instability of TeV pair beams in the IGM over most of the relevant phase space. Oblique instability modes with very small angle $\theta$ relative to the TeV pair beam survive because the linear Landau damping by the cosmic ray background is driven by particles with velocity $v\ge c\cos\theta$, or equivalently Lorentz factors $\gamma>1/\sin\theta$.

% ADD: While instabilities may not quench electromagnetic cascade, they could still have some significance of measuring IGMFs.
\changetext{Pair beam instabilities can lead to appreciable deviations from the standard model of electromagnetic cascade development from TeV sources, which are a key probe of the intergalactic magnetic field \citep{2021Univ....7..223A}. \cite{Broderick_2018} discusses the absence of $\gamma$-ray halos around TeV blazars, suggesting that novel physical processes, such as plasma instabilities, are at play. The difficulty of explaining the absence of secondary GeV photons, either in halos or as a diffuse background, without plasma instabilities has grown in recent years \citep{2023arXiv230301524B}. This underscores the need for a more comprehensive understanding of plasma instabilities and the conditions for them to operate; our findings dramatically narrow, but do not eliminate, the space for a successful plasma instability.}
%to refine our models of IGMF, particularly in light of TeV gamma rays' role as probes of cosmic magnetism.}

Our findings are subject to model uncertainties, but by far the largest is the gamma ray background model. In this paper, we adopt the model in \citet{2019MNRAS.484.4174K} as our foundational EBL source model. 
The fractional contribution of cosmic ray electrons to the IGM pressure at mean density using the alternative \citet{2012ApJ...746..125H} model is 0.5\% at $z = 0.1$, 0.22\% at $z = 1$, and 0.05\% at $z = 2$; this is a factor of 3.6, 4.5, and 6 lower than in the \citet{2019MNRAS.484.4174K} model, respectively.
The \citet{2012ApJ...746..125H} model was intended primarily for ultraviolet/X-ray applications, and it exhibits an exponential cutoff in the energy distribution for Type 2 AGN instead of matching observed $\ge$MeV gamma ray background.
The \citet{2019MNRAS.484.4174K} model, on the other hand, extends the Type 2 AGN spectrum into the MeV region and normalizes it to match the observed background. This is not the only possible source for the gamma-ray background, but as long as the source invoked to explain it exhibits cosmological evolution reasonably similar to star formation or AGN activity, the specific model choice becomes less critical for estimating a background model. Therefore, we believe the use of the \citet{2019MNRAS.484.4174K} model as our fiducial model is appropriate.
However, the linear Landau damping rate is so enormous that even order-of magnitude revisions in the MeV gamma ray background or the redshift dependence of its emissivity would not change the conclusion.

The presence of cosmic ray electrons everywhere in the IGM has several other potential implications. One is on the Lyman-$\alpha$ forest, where extremely high precision measurements are now possible and the interpretation might be affected by the small non-thermal contributions to the pressure discussed in this work.
For example, the extended Baryon Oscillation Spectroscopic Survey (eBOSS) has measured the amplitude of fluctuations $\sigma_8$ to $\pm 2.6\%$ \citep{2019JCAP...07..017C}; the ongoing Dark Energy Spectroscopic Instrument (DESI) survey aims to measure the slope of the spectral index $n_s$ to $\pm 0.0029$ via the Lyman-$\alpha$ forest \citep{2016arXiv161100036D}, and future surveys with many more sightlines are being proposed \citep{2022arXiv220904322S}. A new contribution to IGM pressure at the percent level might therefore be significant for the future experiments. On the other hand, there are likely some (partial) degeneracies between the non-thermal pressure and other IGM thermodynamics parameters that are already marginalized in Lyman-$\alpha$ forest analyses, and cosmological parameters such as $\sigma_8$ and $n_s$ can sometimes be constrained quite well even when there are degeneracies in the IGM thermodynamics parameters \citep{2019JCAP...07..017C}. It also ``helps'' that the cosmic ray electron contribution increases toward lower redshift, whereas ground-based Lyman-$\alpha$ forest observations are restricted to $z\gtrsim 1.7$ by atmospheric opacity.
 
Another possible implication is that cosmic rays are implicated in several of the proposed mechanisms for generating magnetic fields in the IGM directly via battery mechanisms \citep{2020ApJ...896L..12O} or via instabilities driven by cosmic ray anisotropy, such as the dipole-driven non-resonant streaming instability, which can grow from a very small but non-zero background field \citep{2004MNRAS.353..550B}; and the quadrupole-driven Weibel-type instabilities, which in the collisionless limit can grow starting from no background magnetic field and an arbitrarily small anisotropy \citep{1959PhRvL...2...83W}, even in the relativistic regime \citep{1969JMP....10...13L}. We have investigated these mechanisms and found them to be likely insufficient for magnetic field generation in the IGM: the order of magnitude of magnetic fields generated via battery mechanisms are estimated to be up to $\sim 10^{-25}$ G, and the timescale for the cosmic ray anisotropy-driven Weibel instability is too long.

In conclusion, although much work remains to understand the implications, it is evident that the cosmic ray electron population is an important ingredient in the evolution of the TeV pair beams from blazars, and more generally in the plasma physics of the intergalactic medium.

\begin{acknowledgments}
We thank John Beacom, Joseph Lazio, Paarmita Pandey, Todd Thompson, and David Weinberg for useful discussions.

During the preparation of this work, the authors were supported by NASA contract 15-WFIRST15-0008, Simons Foundation  grant 256298, David \& Lucile Packard Foundation grant 2021-72096. H.L. was also supported by an Ohio State University Presidential Fellowship.
\end{acknowledgments}

\software{
numpy \citep{2020Natur.585..357H},
{\sc Matplotlib} \citep{Hunter:2007}
}

\appendix

\section{Relativistic Landau damping in a magnetic field}
\label{app:rlm}

\changetext{This appendix computes the rate of cosmic ray-induced Landau damping for oscillations in a weakly magnetized plasma ($\omega_{\rm c}\ll \omega_{\rm p}$). For simplicity, we will focus on the case of a cold plasma with a small population of isotropic cosmic rays. We start with the formulation of the dispersion relation in \cite{2010PhPl...17l0501B}: $\det{\bf T}=0$, where the $3\times 3$ matrix ${\bf T}$ is
\begin{equation}
T_{ij} = \frac{\omega^2}{c^2}\epsilon_{ij}({\boldsymbol k},\omega) + k_ik_j - k^2\delta_{ij}.
\end{equation}
The matrix ${\bf T}$ is defined for real ${\boldsymbol k}$ and (in general) complex $\omega$. If the problem around which we are perturbing has no resonant particles (e.g., the cold plasma, and $|\omega|\neq\omega_{\rm c}$) then ${\bf T}$ is Hermitian for real $\omega$, and so perturbing the eigenvalue equation $\sum_j T_{ij}\zeta_j = 0$ and multiplying by $\zeta^\ast_i$ gives a first-order frequency shift
\begin{equation}
\Delta \omega = -\frac{  \omega^2 \sum_{ij} \zeta_i^\ast \zeta_j \Delta\epsilon_{ij}({\boldsymbol k},\omega)
}{\sum_{ij}
\zeta_i^\ast \zeta_j \partial_\omega [\omega^2\epsilon_{ij}({\boldsymbol k},\omega) ]
},
\label{eq:a-domega}
\end{equation}
where $\Delta\epsilon_{ij}$ is the new contribution to the dielectric tensor.}

\subsection{The cold background}

\changetext{In this case, perturbation theory is applied around the unperturbed case of a cold magnetized plasma, and the ``$\Delta\epsilon_{ij}$'' is the change in dielectric tensor due to the cosmic rays. 
We choose a coordinate system with the magnetic field on the $\bar z$-axis (we use barred indices to indicate the coordinate system aligned with the field and with the wave vector ${\boldsymbol k}$ in the $\bar x\bar z$-plane). Then the dielectric tensor of the cold plasma component is
\begin{equation}
\omega^2\epsilon_{\bar i\bar j} = \left( \begin{array}{ccc}
\omega^2 - \frac1{1-\omega_{\rm c}^2/\omega^2} \omega_{\rm p}^2 & i \frac{\omega_{\rm c}/\omega}{1-\omega_{\rm c}^2/\omega^2} \omega_{\rm p}^2 & 0 \\
-i \frac{\omega_{\rm c}/\omega}{1-\omega_{\rm c}^2/\omega^2} \omega_{\rm p}^2 & \omega^2 - \frac1{1-\omega_{\rm c}^2/\omega^2} \omega_{\rm p}^2 & 0 \\
0 & 0 & \omega^2 - \omega_{\rm p}^2 \\
\end{array}\right)
\label{eq:ooe}
\end{equation}
\citep[e.g.][\S3.4.1]{1993ppic.conf...55D}.
% xddot = (q ydot B + q Ex)/m
% yddot = -(q xdot B + q Ey)/m
% (x+iy)ddot = { -iq (x+iy)dot B + q(Ex+iEy)} / m
% -omega^2 (x+iy) = -q/m * omega (x+iy) B + q/m (Ex+iEy)
% (x+iy) = q/m (Ex+iEy) / {-omega^2 - qB/m * omega}
%
% --> omega_p^2 / (omega^2 + q B/m omega)
% omega_p^2 / (1 + omega_c /omega)
%
% exx ± i eyx = omega_p^2 / (1 + omega_c /omega) ± omega_p^2 / (1 - omega_c /omega)
% exy ± i eyy = [ omega_p^2 / (1 + omega_c /omega) –+ omega_p^2 / (1 - omega_c /omega) ] / i
%
Now if the wave vector is at an angle $\Theta$ to the magnetic field, then $k_{\bar i} = (k\sin\Theta, 0, k\cos\Theta)$. The ${\bf T}$-matrix is:
\begin{equation}
{\bf T} = \frac1{c^2} \left( \begin{array}{ccc}
\omega^2 - \frac1{1-\omega_{\rm c}^2/\omega^2} \omega_{\rm p}^2 -c^2k^2 \cos^2\Theta & i \frac{\omega_{\rm c}/\omega}{1-\omega_{\rm c}^2/\omega^2} \omega_{\rm p}^2 & c^2k^2\sin\Theta\cos\Theta \\
-i \frac{\omega_{\rm c}/\omega}{1-\omega_{\rm c}^2/\omega^2} \omega_{\rm p}^2 & \omega^2 - \frac1{1-\omega_{\rm c}^2/\omega^2} \omega_{\rm p}^2 - c^2k^2 & 0 \\
c^2k^2\sin\Theta\cos\Theta & 0 & \omega^2 - \omega_{\rm p}^2 - c^2k^2 \sin^2\Theta \\
\end{array}\right).
\end{equation}
For a zero eigenvalue, we must have (using the second two rows)
\begin{equation}
\zeta_{\bar y} = i \frac{\omega_{\rm c} \omega_{\rm p}^2 \omega}{(\omega^2 - c^2k^2)(\omega^2-\omega_{\rm c}^2) - \omega_{\rm p}^2\omega^2} \zeta_{\bar x}
~~~{\rm and}~~~
\zeta_{\bar z} = -\frac{c^2k^2\sin\Theta\cos\Theta}{\omega^2 - \omega_{\rm p}^2 - c^2k^2\sin^2\Theta} \zeta_{\bar x},
\end{equation}
which leads to the dispersion relation
\begin{equation}
\omega^2 - \frac1{1-\omega_{\rm c}^2/\omega^2} \omega_{\rm p}^2 -c^2k^2 \cos^2\Theta
-  \frac{1}{1-\omega_{\rm c}^2/\omega^2} \frac{ \omega_{\rm c}^2 \omega_{\rm p}^4}{(\omega^2 - c^2k^2)(\omega^2-\omega_{\rm c}^2) - \omega_{\rm p}^2\omega^2}
- \frac{c^4k^4\sin^2\Theta\cos^2\Theta}{\omega^2 - \omega_{\rm p}^2 - c^2k^2\sin^2\Theta} = 0.
\label{eq:b-disp}
\end{equation}
It is exactly true that for real ${\boldsymbol k}$ and small magnetic field, there is a real solution for $\omega$ (that goes to $\omega_{\rm p}$ as $B\rightarrow 0$), and the eigenvector has $\zeta_{\bar y}/\zeta_{\bar x}$ purely imaginary and $\zeta_{\bar z}/\zeta_{\bar x}$ real: that is, $E_{\bar x}$ and $E_{\bar z}$ are in phase, but $E_{\bar y}$ is $90^\circ$ out of phase.}

\changetext{The denominator in Eq.~(\ref{eq:a-domega}) depends on the derivative of Eq.~(\ref{eq:ooe}) with respect to frequency. This is
\begin{equation}
D \equiv \sum_{ij}
\zeta_i^\ast \zeta_j \partial_\omega [\omega^2\epsilon_{ij}({\bf k},\omega)]
= 2\omega |{\boldsymbol \zeta}|^2 + \frac{\omega_{\rm p}^2\omega_{\rm c}^2}{\omega^3(1-\omega_{\rm c}^2/\omega^2)^2}  \left[ 2(|\zeta_{\bar x}|^2 + |\zeta_{\bar y}|^2)
- i \frac{\zeta_{\bar x}^\ast\zeta_{\bar y} - \zeta_{\bar y}^\ast\zeta_{\bar x}}{\omega_{\rm c}}
\omega \left(1 + \frac{\omega_{\rm c}^2}{\omega^2}\right) \right]
\label{eq:-den}
\end{equation}}
% der x/(1-x^2) = [ 1-x^2 - x(-2x) ]/(1-x^2)^2 = (1+x^2)/(1-x^2)^2

\changetext{Note that in the limit of $\omega_{\rm c}\rightarrow 0$, we have the simple limits of $\zeta_{\bar i}\rightarrow (\sin\Theta,0,\cos\Theta)$ and $D\rightarrow 2\omega_{\rm p}$.}

\subsection{Cosmic ray contribution to the dielectric tensor}

\changetext{We start from the standard formulation for a hot collisionless electron gas to the dielectric tensor. \citet[Eq.~2.9]{Trubnikov59} gives the full derivation starting from the relativistic Vlasov equation, although we use a general form for $\partial f_0^{\rm CR}/\partial p$ rather than assuming a thermal distribution. We write the phase space density written in spherical coordinates $p_\parallel = p\cos \alpha$ and $p_\perp = p\sin \alpha$ (where $\alpha$ is the pitch angle), and with the phase space density $f_0$ set to integrate to $\int f_0^{\rm CR}\,d^3{\bf p} = n_{\rm CR}$. We further assume a spherically symmetric particle distribution, so $\partial f_0^{\rm CR}/\partial\alpha = 0$. The contribution of cosmic rays to the dielectric tensor is then
\begin{equation}
\omega^2 \Delta\epsilon_{\bar i\bar j}^{\rm CR} = 8\pi^2e^2\omega \int_0^\infty p^2\,dp \int_{-1}^1 d(\cos\alpha) \, \sum_{n=-\infty}^\infty \frac{ v\sin^2\alpha}{\omega - k_\parallel v \cos\alpha - n\omega_{\rm s}} \frac{\partial f_0^{\rm CR}}{\partial p} {\mathbb T}_{n,ij}(z),
\end{equation}
where $\omega_{\rm s} = \omega_{\rm c}/\gamma$ and $\omega_{\rm c} = eB/mc$; and
\begin{equation}
{\mathbb T}_n(z) = \left( \begin{array}{ccc}
n^2z^{-2}[J_n(z)]^2 & -inz^{-1}J_n(z)J'_n(z) & nz^{-1}[J_n(z)]^2\cot\alpha \\
inz^{-1}J_n(z)J'_n(z) & [J'_n(z)]^2 & iJ_n(z)J'_n(z)\cot\alpha \\
nz^{-1}[J_n(z)]^2\cot\alpha & -iJ_n(z)J'_n(z)\cot\alpha & [J_n(z)]^2\cot^2\alpha
\end{array} \right)
\end{equation}
with $z = k_\perp v\sin\alpha/\omega_{\rm s}$.
The part we care about is the projection onto the electric field eigenvector. Using the phase convention that $\zeta_{\bar x}$ is real (so $\zeta_{\bar z}$ is also real and $\zeta_{\bar y}$ is purely imaginary), we find
\begin{eqnarray}
\omega^2 \sum_{ij} \zeta_i^\ast \zeta_j \Delta\epsilon_{\bar i\bar j}^{\rm CR} &=& 8\pi^2e^2\omega \int_0^\infty p^2\,dp \int_{-1}^1 d(\cos\alpha) \, \sum_{n=-\infty}^\infty \frac{v}{\omega - k_\parallel v \cos\alpha - n\omega_{\rm s}} \frac{\partial f_0^{\rm CR}}{\partial p}
\nonumber \\ && \times
\Biggl[
 \frac nzJ_n(z)\zeta_{\bar x} \sin\alpha +  J_n(z) \zeta_{\bar z} \cos\alpha +  J'_n(z) \frac{\zeta_{\bar y}}i \sin\alpha \Biggr]^2.
\label{eq:cr-diel2}
\end{eqnarray}
It can be seen that under most circumstances, the integrand in Eq.~(\ref{eq:cr-diel2}) is real (recall that $\zeta_{\bar x}$ and $\zeta_{\bar z}$ are real but $\zeta_{\bar y}$ is purely imaginary. The exception is when a denominator goes to zero, since then we must detour around the singularity. In the event of a singularity, we apply an infinitesimal positive imaginary part to $\Im \omega$ (corresponding to a perturbation that vanishes in the distant past). Using the rule that
\begin{equation}
\frac1{\omega-\omega_0} \rightarrow {\mathbb P} \frac1{\omega-\omega_0} - i \pi \delta(\omega-\omega_0),
\end{equation}
we find that in the imaginary part we can collapse the integral over $\cos\alpha$, leaving behind only the sum:
\begin{eqnarray}
\Im [\omega^2 \sum_{ij} \zeta_i^\ast \zeta_j \Delta\epsilon_{\bar i\bar j}^{\rm CR}] &=& -8\pi^3e^2\omega \int_0^\infty p^2\,dp \,\frac{\partial f_0^{\rm CR}}{\partial p} \frac{v\,\Delta\mu}{\omega_{\rm s}} \sum_n 
\nonumber \\ && \times
\Biggl[
 \frac nzJ_n(z)\zeta_{\bar x} \sin\alpha +  J_n(z) \zeta_{\bar z} \cos\alpha +  J'_n(z) \frac{\zeta_{\bar y}}i \sin\alpha \Biggr]^2,
\label{eq:cr-diel3}
\end{eqnarray}
where $n$ is an integer, $\cos\alpha = (\omega-n\omega_{\rm s})/k_\parallel v$, the range of the sum $n$ is over the allowed values where $-1\le\cos\alpha\le 1$, and the spacing of $\mu=\cos\alpha$ values is $\Delta\mu = \omega_{\rm s}/|k_\parallel|v$. The growth rate of a plasma oscillation is then
\begin{equation}
\Im\omega = \frac{2\pi^2 \omega_{\rm p}^2\omega}{Dn_{e,\rm th}} \int_0^\infty p^2\,dp \,\frac{\partial f_0^{\rm CR}}{\partial p} \frac{m_ev\,\Delta\mu}{\omega_{\rm s}} \sum_n 
\Biggl[
 \frac nzJ_n(z)\zeta_{\bar x} \sin\alpha +  J_n(z) \zeta_{\bar z} \cos\alpha +  J'_n(z) \frac{\zeta_{\bar y}}i \sin\alpha \Biggr]^2.
\label{eq:growth-mag}
\end{equation}
As usual, a negative $\Im\omega$ indicates damping.}

\bibliography{main}{}
\bibliographystyle{aasjournal}

\end{document}